\documentclass[onecolumn,amsmath,amssymb,11pt,superscriptaddress,nofootinbib]{revtex4}

\usepackage{graphicx}
\usepackage{amsmath,amssymb,amsfonts,amsthm,stmaryrd,mathtools,bm,physics,tensor}
\usepackage{color}
\usepackage{tikz}
\allowdisplaybreaks[1]

\usepackage[bookmarks,linktocpage, colorlinks=true, plainpages = false, citecolor = treegreen,  linkcolor=darkblue, urlcolor = darkblue, filecolor = blue]{hyperref} 

\def\be#1\ee{\begin{align}#1\end{align}}

\def\ba{\begin{eqnarray}}
	\def\ea{\end{eqnarray}}
\def\nn{\nonumber}

\definecolor{tealgreen}{rgb}{0.0, 0.5, 1.0}
\definecolor{darkblue}{rgb}{0., 0.4, 0.8}
\definecolor{cadmiumred}{rgb}{1., 0., 0.22}
\definecolor{treegreen}{rgb}{0., 0.7, 0.3}

\begin{document}

\title{Quantum dynamics and thermodynamics of a Minkowski-Minkowski wormhole
}

\author{Johanna Borissova}
\email{j.borissova@imperial.ac.uk}
\affiliation{Abdus Salam Centre for Theoretical Physics, Imperial College London, London SW7 2AZ, UK}
\author{Jo\~ao Magueijo}
\email{j.magueijo@imperial.ac.uk}
\affiliation{Abdus Salam Centre for Theoretical Physics, Imperial College London, London SW7 2AZ, UK}

\begin{abstract}
\bigskip
{\sc Abstract:} We consider the path-integral quantization of a minisuperspace cut-and-paste Lorentzian wormhole connecting two Minkowski spacetimes. The dynamics of the throat radius as a function of proper time is governed by a non-polynomial effective action derived by an application of the Israel junction condition formalism. Within a saddle-point approximation of the propagator describing the evolution from an initial to a final throat radius, we show that topology-changing transitions are suppressed by the Hessian determinant. In addition, we analyze the gravitational thermodynamics of the wormhole spacetime by a Wick rotation of the Israel-Lanczos equations in the presence of a thin-shell source. The resulting Euclideanized field equations are assumed to originate from a Euclidean effective gravity-matter action, which enters the path-integral representation of the gravitational canonical partition function. Therefrom we associate a temperature given by the inverse period of  solutions, as well as a gravitational entropy as functions of the surface energy density and equation of state parameter of the shell. Both quantities are sourced entirely by the discontinuity of the extrinsic curvature across the junction. We show how this result can be applied to deduce a thermodynamic first law as the differential version of the conservation equation relating the effective mass of the shell to its surface pressure. 

\end{abstract}

\maketitle
\tableofcontents

\section{Introduction}\label{Sec:Introduction}

Traversable wormholes are fascinating constructs that enable the hypothetical journey from one universe to another. Initially popularized by Morris and Thorne~\cite{Morris:1988cz,Morris:1988tu}, they have become objects of considerable scientific interest across the classical and quantum-gravity community~\cite{Buoninfante:2024yth,Carballo-Rubio:2025fnc}. Two primary areas in which wormholes play a central role are the question of topology change in quantum gravity~\cite{Geroch:1967fs,Horowitz:1991fr,Anderson:1986ww,Louko:1995jw,Dowker:2002hm} and the long-standing debate about the fate of global charges in our universe~\cite{Banks:1988yz,Giddings:1987cg,Lee:1988ge,Abbott:1989jw,Coleman:1989zu}. The answers to these questions can be expected to depend substantially on the specifics of the quantum-gravity theory~\cite{Borissova:2024hkc,Basile:2025zjc}. 

Different from standard classical black holes, in which geodesics converge until they terminate at a spacetime singularity, a wormhole geometry is free from central singularities. Achieving the necessary defocusing of null rays requires a violation of the null convergence condition near the throat~\cite{Morris:1988cz,Morris:1988tu,Visser:1995cc,Hochberg:1998ii,Simpson:2019cer,Simpson:2018tsi,Borissova:2025msp}. When the dynamics is governed by the Einstein equations, the latter becomes equivalent to the null  energy condition, which for a perfect fluid states that the sum of its energy density and pressure must be non-negative. This fact is often rephrased as the statement that wormholes in the theory of general relativity can only be sourced by exotic type of matter with negative energy density, or negative pressure, or both. In fact, more generally a violation of the null convergence condition necessarily implies a violation of the timelike convergence condition, see e.g.~\cite{Borissova:2025hmj}.

There exist many distinct proposals to construct wormhole geometries. Here, we will focus on a class of Lorentzian cut-and-paste traversable wormholes  introduced originally by Visser~\cite{Visser:1989am,Visser:1989kg,Visser:1989kh,Visser:1990wh,Visser:1990wi,Visser:1995cc}. These are obtained by cutting out regions from two given spacetimes and subsequently gluing them along a timelike junction surface which thereby becomes the throat of the wormhole. An effective action governing the dynamics of the throat as a function of proper time can be obtained by evaluating the Einstein action for the wormhole spacetime. The latter contains non-polynomial  contributions due to the delta-function singularity of the Riemann curvature on the throat, which is reminiscent of lower-dimensional crotch singularities encountered in the discussion of topology change via the trousers or yarmulke cobordisms~\cite{Louko:1995jw,Anderson:1986ww,Dowker:2002hm}. 

In~\cite{Visser:1989am,Visser:1990wh,Visser:1990wi,Visser:1995cc} such types of dynamical wormhole geometries have been quantized canonically by solving a Wheeler-de-Witt type equation $\hat{H}\psi = 0$, where the non-local Hamiltonian $H$ derived from the wormhole effective action appears in one of the Israel-Lanczos equations. The considered wormholes have been found to be quantum mechanically stable with an average radius of the order of one Planck length, thereby leading to the conclusion that topology change is suppressed by quantum effects. 

Our main goal in the following is to reconsider such types of questions from the point of view of a minisuperspace Lorentzian gravitational path integral defined on the configuration space of wormhole throat radial functions, whose dynamics is governed by the aforementioned effective action. The putative constraint relation $H=0$ representing one of the Israel-Lanczos equations, which has been quantized in the works~\cite{Visser:1989am,Visser:1990wh,Visser:1990wi,Visser:1995cc}, is not imposed by effective action itself, which in fact is not reparametrization invariant. Instead, this relation arises as an integrated equation of motion. This suggests that in a path-integral treatment of the wormhole dynamics, one should perform a non-relativistic path integral on the reduced configuration space of throat radial functions only, as opposed to a relativistic path integral in which the constraint of a vanishing Hamiltonian would have to be dealt with separately through an integral over a Lagrange multiplier playing the role of a lapse, see e.g.~\cite{Teitelboim:1981ua,Teitelboim:1983fh,Teitelboim:1983fk,Hartle:1986yu,Halliwell:1988ik,Halliwell:1988wc,Banihashemi:2024aal}. A similar viewpoint has been adopted also in~\cite{Redmount:1993ue}.

In addition, we will associate a wormhole gravitational thermodynamics from the perspective of a Euclidean path integral over radial functions periodic in imaginary time, with a Euclidean effective gravity-matter action assumed to give rise to the Wick-rotated Israel-Lanczos equations in the presence of a thin-shell source. We shall see that the thereby obtained gravitational temperature and entropy are constants sourced entirely by the discontinuity of the extrinsic curvature across the junction, with a dependence of the zero-point surface energy density to a power characteristic to the equation of state of the mass shell. Moreover, we shall see that the wormhole gravitational entropy as a function of temperature behaves as
\ba
S \,\,\,\sim \,\,\, \frac{1}{T^2}\,.
\ea
Such a scaling is characteristic of gravitational systems with horizons, such as black-hole geometries and de Sitter space.~\footnote{Note that the temperature and entropy are a priori independent degrees of freedom in geometrical theories.} Remarkably, the same relation is recovered here in the absence of a horizon and in the presence of a timelike junction surface. This phenomenon is not entirely new: analogous behaviour arises, e.g., in the Giddings–Strominger wormhole in the context of axion-induced gravity~\cite{Giddings:1987cg,Giddings:1988cx}. Further analogies may be drawn with the spacetimes containing timelike junctions analyzed in~\cite{Isichei:2025tvg}, even in the context of the induced matter thermodynamics. These observations suggest that features often attributed to spacetimes with horizons may, in fact, be more general properties of gravitational systems.

This paper is structured as follows. In Section~\ref{Sec:MinkowskiMinkowskiWormhole} we describe the cut-and-paste procedure leading to a wormhole which  connects two Minkowski spacetimes. In Section~\ref{Sec:WormholeAction} we review the derivation of the effective action describing the dynamics of the throat radius by an application of the Israel junction condition formalism. In Section~\ref{Sec:PathIntegralWKB} we consider a WKB approximation for the propagator expressed as a path integral over throat radial functions. In Section~\ref{Sec:Thermodynamics} we consider the thermodynamics arising from a path-integral representation of the canonical  partition function, with a Euclidean effective action giving rise to the Wick-rotated Israel-Lanczos equations in the presence of a thin matter shell.  We finish with a dicussion in Section~\ref{Sec:Discussion}.

\section{Classical Minkowski-Minkowski wormhole}\label{Sec:MinkowskiMinkowskiWormhole}

We will focus on a simple model of a Minkowskian cut-and-paste Lorentzian wormhole~\cite{Visser:1989am,Visser:1989kg,Visser:1989kh,Visser:1990wh,Visser:1990wi,Visser:1995cc}. Let $M_\pm$ denote two copies of the Minkowski spacetime. The Minkowski line element in spherical coordinates $x^\mu = (t,r,\theta,\phi)$ is given by
\ba\label{eq:MinkowskiMetric}
\dd{s}^2 = g_{\mu\nu}\dd{x}^\mu \dd{x}^\nu = -\dd{t}^2 +\dd{r}^2 + r^2 \dd{\Omega}^2\,,
\ea
where $\dd{\Omega}^2$ denotes the area element of the unit $2$-sphere.
Each copy $M_\pm$ has its distinct coordinate chart $x^\mu_\pm$, but we will omit the subscript unless necessary.

A traversable wormhole can be obtained as follows. First,  two identical $4$-dimensional spacetime regions
\ba
\Omega_\pm  = \qty{x^\mu = (t,r_\pm,\theta,\phi)\, | \,r_\pm \leq a}
\ea
 are removed from each of the two  Minkowski spacetimes. Subsequently, the resulting geodesically incomplete spacetimes $\mathcal{M}_\pm = M_\pm\setminus \Omega_i$ are identified along their boundaries
 \ba
 \partial \Omega_\pm = \qty{x^\mu = (t,r_\pm,\theta,\phi) \,|\, r_\pm = a}\,.
 \ea
  Thereby a geodescially complete manifold $\mathcal{M}=\mathcal{M}_+ \oplus \mathcal{M}_-$ is obtained, without boundary and with two asymptotically flat regions connected through a wormhole. The wormhole throat is defined as the hypersurface 
  \ba
  \Sigma = \partial\Omega_+ = \partial\Omega_-\,.
  \ea
  Making $\Sigma$ dynamical yields a model for a wormhole with perturbations of the radial location of the throat.  To that end, the constant radius $a$ is promoted to a function depending on the proper time $\tau$ measured by an observer on the throat. The replacement $a \mapsto a(\tau)$ in the intrinsic metric $\gamma_{ij}$ 
on $\Sigma$ leads to
\ba\label{eq:IntrinsicMetric}
\dd{s_{\Sigma}}^2  = \gamma_{ij}\dd{\xi}^i \dd{\xi}^j = -\dd{\tau}^2 + a(\tau)^{2}\dd{\Omega}^2\,.
\ea
Here $\xi^i = (\tau,\theta,\phi)$ with  $i=1,2,3$ denote $3$-dimensional coordinates on $\Sigma$.

\section{Wormhole effective action}\label{Sec:WormholeAction}

In the following we will review the derivation of an effective minisuperspace action for the wormhole on the reduced configuration space of throat radial functions. To that end, we will evaluate the Einstein action as originally considered in~\cite{Visser:1989am}. For completeness, in Appendix~\ref{App:ReparametrizationInvariance} we moreover illustrate how reparametrization invariance of the action can be regained on the extended configuration space. Subsequently, we will proceed to evaluate the non-relativistic path integral for the reduced system. The analogue treatment for the non-relativistic particle is reviewed in Appendix~\ref{App:NonRelativsticParticle}.

The wormhole spacetime $\mathcal{M}$ is determined by the flat Minkowski background everywhere except on the throat $\Sigma $. The scalar curvature on $\Sigma$ can be computed by an application of the junction condition formalism.
We will primarily follow the derivation in~\cite{Berry:2020tky,Visser:1989am,Visser:1989kg,Visser:1990wi,Visser:1995cc}.  The  condition $\partial\Omega_+ = \partial\Omega_-$ identifies the $3$-dimensional geometries at the junction and  ensures that the metric is continuous. The metric is however not differentiable at the junction. Therefore the affine connection is discontinous and produces a delta-function singularity of the Riemann tensor on $\Sigma$. The latter can be expressed in terms of the discontinuity  
\ba\label{eq:DiscontinuityK}
\Delta {K_\Sigma}\indices{^i_j} =  {K_{+}}\indices{^i_j}-{K_{-}}\indices{^i_j}
\ea
of the extrinsic curvature of $\Sigma$ in transiting from $\mathcal{M}_+$ to $\mathcal{M}_-$.
The location of $\Sigma$ with respect to the background $\mathcal{M}$ is described by the $4$-vector $X^\mu(\xi^i)  = \qty(t(\tau),a(\tau),\theta,\phi) $, or equivalently by the roots of the defining function 
\ba
f_\Sigma\qty(x^\mu(\xi^i)) = r - a(\tau)\,,
\ea 
where $x^\mu = (t,r,\theta,\phi)$ denote the $4$-dimensional coordinates on $\mathcal{M}$. Thus $e^\mu_{(i)}= \derivative{X^\mu}{\xi^i}$ define three linearly independent vectors tangent to $\Sigma$. 
The definition of the extrinsic curvature requires the unit normal vectors ${n_\Sigma^\pm}^{\mu}$ to $\Sigma$. These can be derived from the conditions ${n_\Sigma^\pm}_\mu e^\mu_{(i)}=0$ for $i=1,2,3$ and ${n_\Sigma^\pm}_\mu {n_\Sigma^\pm}^{\mu }= +1$, or alternatively by the formula
\ba\label{eq:Normal}
{n_\Sigma^\pm}_\mu = \pm \abs{g^{\alpha\beta}\pdv{f_\Sigma}{x^\alpha}\pdv{f_\Sigma}{x^\beta}}^{-\frac{1}{2}}\pdv{f_\Sigma}{x^\mu}
= \pm\qty(\dot{a},\sqrt{1+\dot{a}^2},0,0)\,,
\ea
where an overdot denotes the derivative with respect to $\tau$.
For a static wormhole ($\dot{a}=0$) the unit normals  reduce to ${n_\Sigma^\pm}\indices{^\mu} \propto \pm(\partial_r)^\mu$. 
The extrinsic curvature of $\Sigma$ can be expressed as
\ba\label{eq:ExtrinsicCurvatureFormula}
{K_\Sigma^\pm}_{ij} =\eval{\nabla^\pm_{(\mu} {n_\Sigma}^\pm_{\nu)} e^\mu_{(i)}e^\nu_{(j)}}_{\Sigma} = - \eval{{n_\Sigma^\pm}_{\mu} \qty(\pdv[2]{X^\mu}{\xi^i}{\xi^j} + \Gamma\indices{^\pm^\mu_\alpha_\beta}\pdv{X^\alpha}{\xi^i}\pdv{X^\beta}{\xi^j} )}_{\Sigma}\,,
\ea
where $\nabla^{\pm}$ and $\Gamma^\pm$ denote the covariant derivatives and Christoffel symbols of the Levi-Civita connections of the $4$-dimensional spacetimes on each side of the throat. In this case, $\Gamma\indices{^\pm^\mu_\alpha_\beta} =  \Gamma\indices{^\mu_\alpha_\beta} $ are the Christoffel symbols of the Minkowski metric~\eqref{eq:MinkowskiMetric} in spherical coordinates.~\footnote{The non-zero independent Christoffel symbols  of the Minkowski metric~\eqref{eq:MinkowskiMetric} in the chart $(t,r,\theta,\phi)$ are given by
\ba
 \Gamma\indices{^r_\theta_\theta} = -r\,,\quad \Gamma\indices{^r_\phi_\phi} = -r\sin^2\theta\,,\quad 
\Gamma\indices{^\theta_r_\theta} &=& \Gamma\indices{^\phi_r_\phi}= \frac{1}{r}\,,\quad
\Gamma\indices{^\theta_\phi_\phi} = -\sin\theta \cos\theta\,,\quad \Gamma\indices{^\phi_\theta_\phi}=\cot\theta\,.
\ea}

 The computation of the extrinsic curvature $ {K_\Sigma^\pm}\indices{^i_j}$ is considerably simplified by making use of the reflection symmetry of the wormhole construction and the spherical symmetry of the Minkowski background. The former implies  $ {K_\Sigma^-}\indices{^i_j} = -  {K_\Sigma^+}\indices{^i_j}$, such that the discontinuity of the extrinsic curvature~\eqref{eq:DiscontinuityK} reduces to $\Delta {K_\Sigma}\indices{^i_j} = 2 {K_\Sigma^{+}}\indices{^i_j} \equiv 2 {K_\Sigma}\indices{^i_j}  $. Spherical symmetry moreover implies that this matrix is diagonal with only two independent components,
 \ba
\qty(\Delta {K_\Sigma}\indices{^i_j}) = 2 \qty({K_\Sigma}\indices{^i_j}) = 2 \,\text{diag}\qty{{K_\Sigma}\indices{^\tau_\tau},{K_\Sigma}\indices{^\theta_\theta},{K_\Sigma}\indices{^\theta_\theta} }\,.
 \ea
  Using~\eqref{eq:ExtrinsicCurvatureFormula}, these are computed as
\ba
{K_\Sigma}\indices{^\theta_\theta} &=&  \frac{\sqrt{1+\dot{a}^2}}{a}\,,\label{eq:ExtrinsicCurvatureComponents1}\\
{K_\Sigma}\indices{^\tau_\tau} &=& \frac{\ddot{a}}{\sqrt{1+\dot{a}^2}} = \derivative{\,\text{arcsinh}(\dot{a})}{\tau}\,.\label{eq:ExtrinsicCurvatureComponents2}
\ea
Denoting by $\eta(x)$ a Gaussian coordinate normal to the throat (with $\eta>0$ in $\mathcal{M}_+$, $\eta<0$ in $\mathcal{M}_-$, and $\eta =0$ on $\Sigma$), the Ricci tensor $R\indices{^\mu_\nu}$ and Ricci scalar $R$ are given by~\footnote{The convention $\Theta(0)\equiv 0$ is used for the Heaviside step function.}
\ba
R\indices{^\mu_\nu}(x)&=&  {R_{+}}\indices{^\mu_\nu}\Theta(\eta) + {R_{-}}\indices{^\mu_\nu}\Theta(-\eta) -\begin{bmatrix}
	\qty(\Delta {K_\Sigma}\indices{^i_j} ) & 0 \\
	0 & \Tr(\Delta K_{\Sigma})
\end{bmatrix}\delta(\eta) \,,\label{eq:RicciTensor}\\
R &=&  R_{+} \Theta(\eta) + R_{-} \Theta(-\eta)-2 \Tr(\Delta K_\Sigma)\delta(\eta) \label{eq:RicciScalar}\,.
\ea
Here $ {R_{\pm}}\indices{^\mu_\nu}$ and $R_{\pm}$ denote the Ricci tensors and Ricci scalars of the two background geometries $\mathcal{M}_\pm$ on both sides of the throat.  The trace of the discontinuity of the extrinsic curvature in transiting from $\mathcal{M}_+$ to  $\mathcal{M}_-$ is given by
\be\label{eq:TraceDiscontinuityExtrinsicCurvature}
\Tr(\Delta K_\Sigma) = 2 \Tr(K_\Sigma) = 2 \qty({K_\Sigma}\indices{^\tau_\tau} + 2 {K_\Sigma}\indices{^\theta_\theta} )\,.
\ee 

The previous formulae allow us to evaluate the Einstein-Hilbert action for the wormhole spacetime $\mathcal{M}$. It takes the form~\footnote{We work in units $G=\hbar = c = 1$.}
\ba\label{eq:ActionDefinition}
S_{\text{wormhole}}
&=& \frac{1}{16\pi}\int_{\mathcal{M}} \dd[4]{x}\sqrt{-g}R = \sum_{i=1,2}  \qty(\frac{1}{16\pi} \int_{\mathcal{M}_i} \dd[4]{x}\sqrt{-g_{(i)}}R_{(i)})  -\frac{1}{8\pi} \int_{\Sigma} \dd[3]{x}\sqrt{-\gamma}\Tr(\Delta K_\Sigma)
\nn\\
&=& S_{\text{background}} + S_{\text{throat}}\,.
\ea
In the above expression
$\gamma$ denotes the determinant of the induced metric~\eqref{eq:IntrinsicMetric} on $\Sigma$. The Ricci scalars $R_{\pm}$ of the  flat Minkowski backgrounds $M_\pm$ vanish. Therefore, after substituting~\eqref{eq:ExtrinsicCurvatureComponents1}--\eqref{eq:ExtrinsicCurvatureComponents2} for the components of the extrinsic curvature, the wormhole action reduces to
\ba\label{eq:ActionEvaluated}
S_{\text{wormhole}}
&=&S_{\text{throat}}=-\frac{1}{8\pi} \int_{\Sigma} \dd[3]{\xi}\sqrt{-\gamma}\Tr(\Delta K_\Sigma) \nn\\
&=& -\frac{1}{8\pi} \int_{0}^{2\pi}\dd{\phi}\int_0^{\pi}\dd{\theta}\sin\theta \int \dd{\tau}a^2 \cdot 2\qty(\derivative{\,\text{arcsinh}(\dot{a})}{\tau} + 2 \frac{\sqrt{1+\dot{a}^2}}{a}) \nn\\
&=& \int  \dd{\tau} 2 \qty( a \dot{a}\,\text{arcsinh}({\dot{a}}) - a \,\sqrt{1+\dot{a}^2}) \equiv \int \dd{\tau} L(a,\dot{a}) \,.
\ea
Here we have integrated by parts to remove the derivative from the first term and dropped the boundary terms. The resulting Lagrangian describes the dynamics of a single degree of freedom given by the throat radius $a(\tau)$ as a function of proper time $\tau $ on $\Sigma$ and has been derived originally in~\cite{Visser:1989am}. It should be noted that this effective Lagrangian is non-polynomial in the velocity $\dot{a}$. The Euler-Lagrange equations derived from this Lagrangian are
\ba
 \derivative{}{\tau}\qty(\pdv{L}{\dot{a}}) - \pdv{L}{a} \,=\,0  \,\,\, \quad \Rightarrow \quad \,\,\, a \ddot{a} + \dot{a}^2 + 1 \,=\, 0\,,\label{eq:ELEOM}
\ea
which is equivalent to
\ba\label{eq:ELEOMRewritten}
\derivative[2]{\qty[a^2]}{\tau} &=& -2\,.
\ea
For later reference we also note that the momentum conjugated to the variable $a$ in~\eqref{eq:ActionEvaluated} is
\ba\label{eq:ConjugateMomentum}
p = \pdv{L}{\dot{a}} = 2 a \,\text{arcsinh}(\dot{a})\,.
\ea
Inverting the previous relation to obtain $\dot{a} = \dot{a}(p,a)$ yields the canonical Hamiltonian \ba\label{eq:Hamiltonian}
H(a,p) = p \dot{a} - L =  2 a \cosh(\frac{p}{2a}) \,\,\,\quad \Leftrightarrow \,\,\,\quad H(a,\dot{a}) = 2 a\sqrt{1+\dot{a}^2} \,.
\ea
 
\section{Path integral and WKB approximation}\label{Sec:PathIntegralWKB}

	\begin{figure}[t]
	\centering
	\includegraphics[width=.46\textwidth]{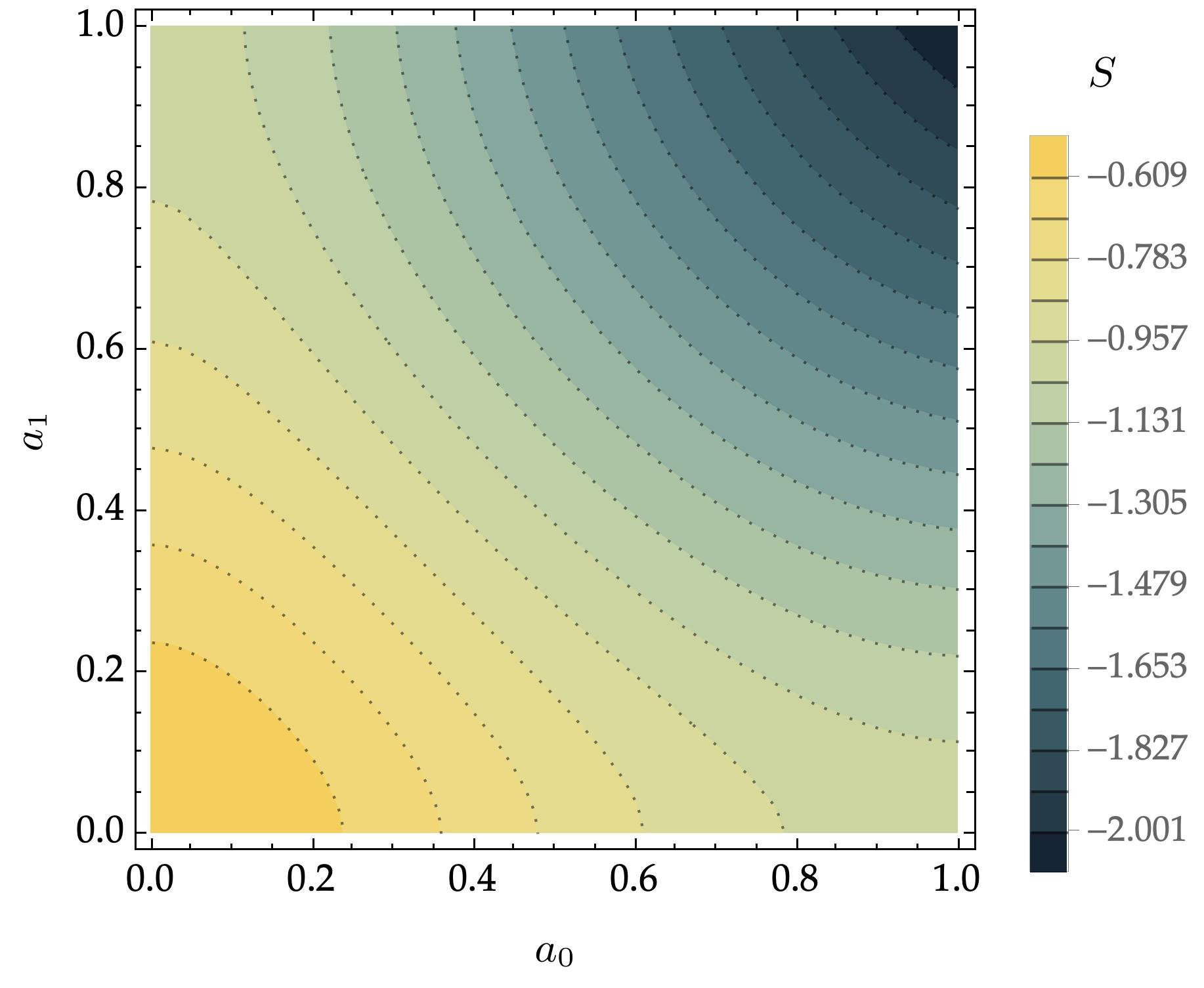}
	\hspace{0.4cm}
	\includegraphics[width=.5\textwidth]{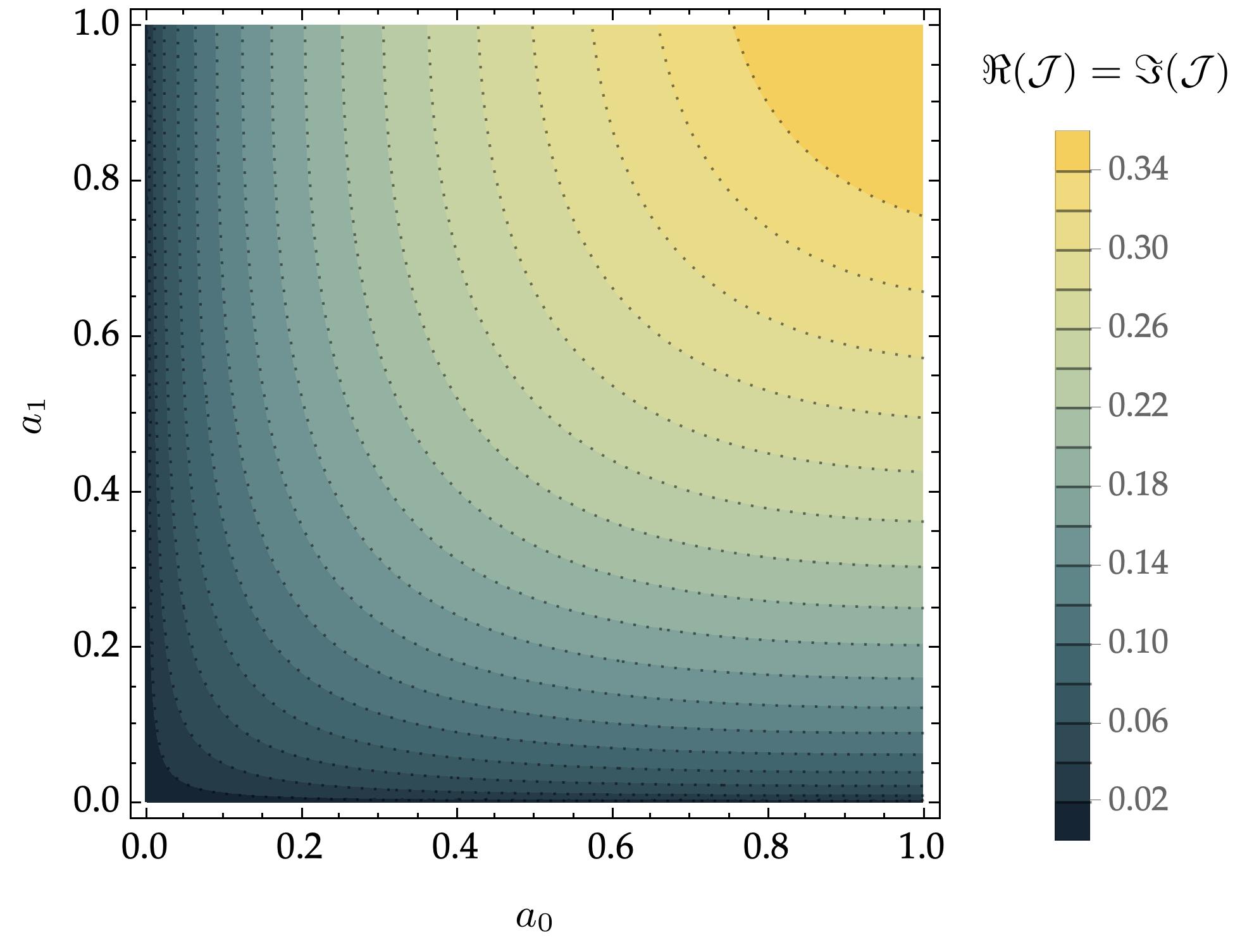}
	\caption{\label{Fig:SJ} Onshell wormhole action $S$ as given in~\eqref{eq:SOnshell} for a choice of proper-time interval $\Delta \tau = 1$, as well as real and imaginary parts of the Hessian factor $\mathcal{J}$ as a function of the initial and final throat radius $a_0$ and $a_1$ for $\Delta \tau = 1$.}
\end{figure} 
 
In the following, starting from the effective action~\eqref{eq:ActionEvaluated}, we consider the path integral
\ba\label{eq:PathIntegral}
	G( a_1,\tau_1; a_0,\tau_0 ) = \int_{a(\tau_0)=a_0}^{a(\tau_1)=a_1} \mathcal{D}a\,e^{\imath S[a(\tau)]}
\ea
defined on the configuration space of throat radial functions $a(\tau)$ satisfying the boundary conditions $a(\tau_0)=a_0$ and $a(\tau_1)=a_1$. This path integral represents the quantum propagator for the evolution of the wormhole from an initial throat radius $a_0$ to a final throat radius $a_1$ within a time interval $\Delta \tau = \tau_1 - \tau_0$. 

		\begin{figure}[t]
	\centering
	\includegraphics[width=.47\textwidth]{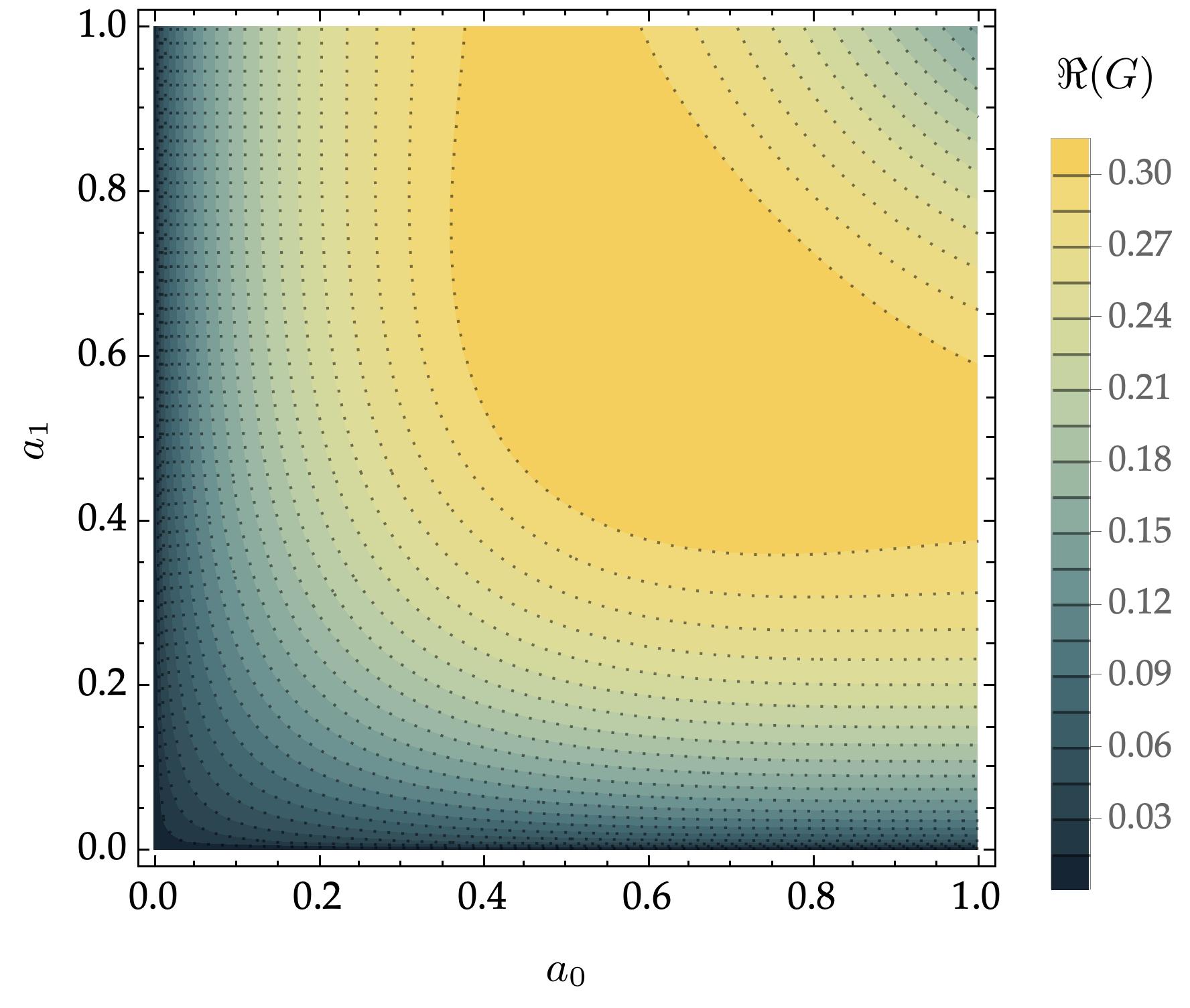}
	\hspace{0.4cm}
	\includegraphics[width=.483\textwidth]{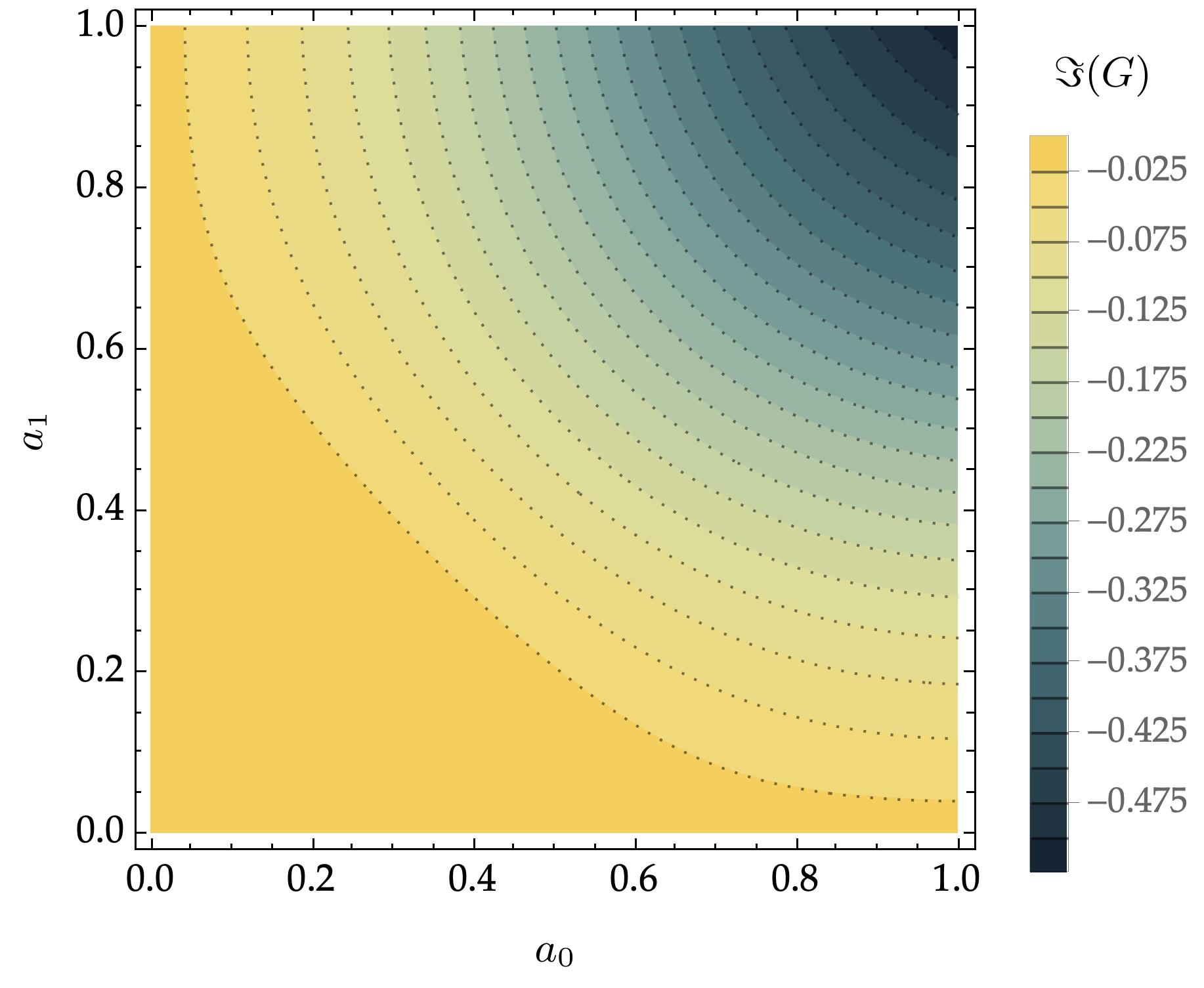}
	\caption{\label{Fig:G} Real and imaginary parts of $G$ as a function of the initial and final throat radius $a_0$ and $a_1$ for $\Delta \tau = 1$}.
\end{figure}

\begin{figure}[t]
	\centering
	\includegraphics[width=.47\textwidth]{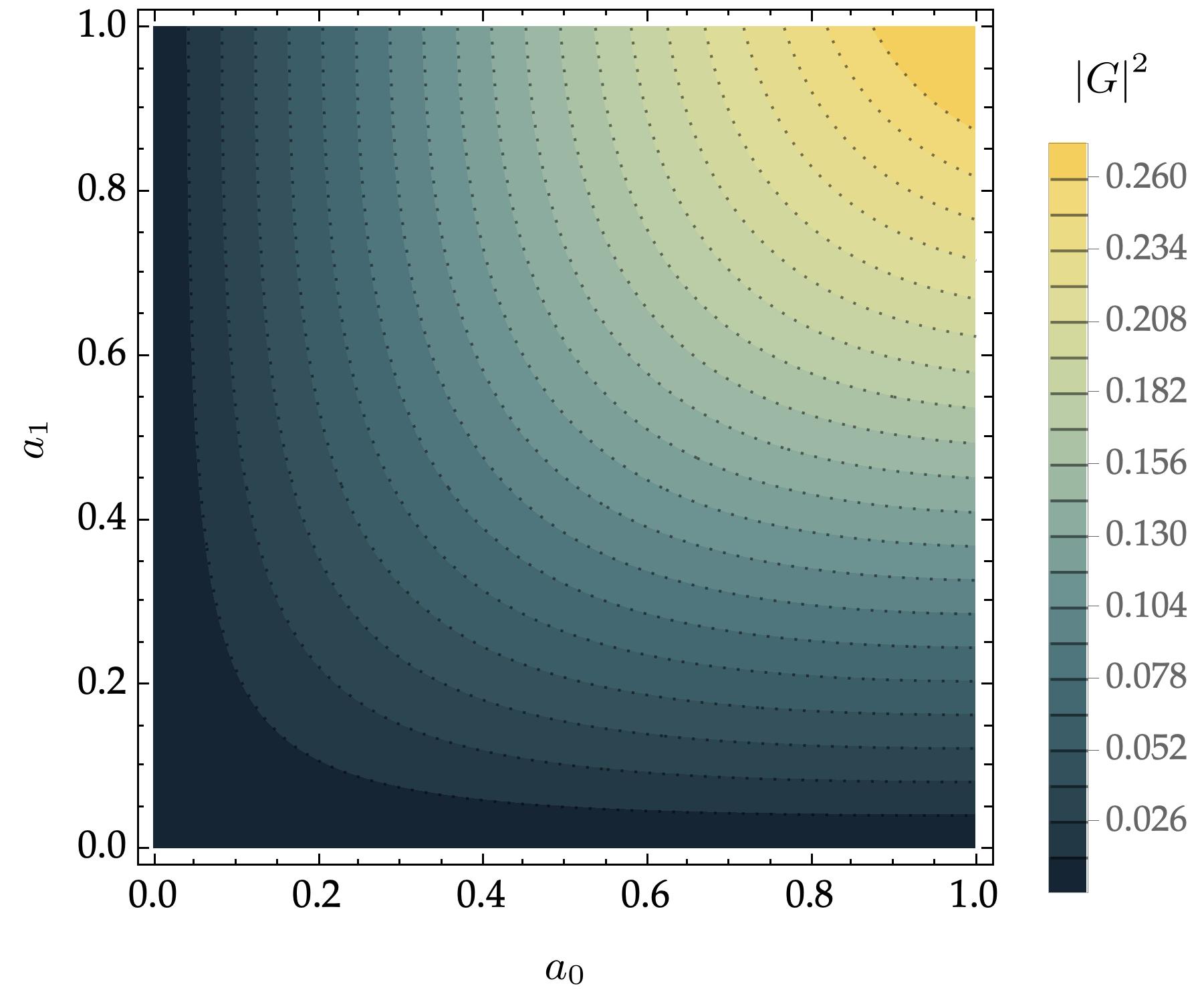}
	\caption{\label{Fig:GGStar} Probability distribution $\abs{G}^2 =GG^*$ as a function of the initial and final throat radius $a_0$ and $a_1$ for $\Delta \tau = 1$.}
\end{figure}

The non-polynomial form of the effective action~\eqref{eq:ActionEvaluated} obstructs an exact evaluation of the propagator. On these grounds we will resort to a Wentzel-Kramers-Brillouin (WKB) analysis and approximate~\eqref{eq:PathIntegral} as 
\ba\label{eq:PathIntegralApproximation}
G(a_1, \tau_1; a_0,\tau_0 ) \, \,\sim\,\,
\sum_{\sigma} \mathcal{J}_\sigma\,
e^{\imath
	S_\sigma}\,\,\,\quad \text{where}\quad  \,\,\, \mathcal{J}_\sigma =  \sqrt{\frac{1}{2 \pi \imath}\pdv[2]{S_\sigma}{a_0}{ a_1} }\,.
\ea
Here $S_\sigma$ denotes the action evaluated on classical solutions  $\sigma$ to the Euler-Lagrange equations~\eqref{eq:ELEOM} satifying the boundary conditions $a(\tau_0) = a_0$ and $a(\tau_1) = a_1$.  Written in the form~\eqref{eq:ELEOMRewritten} and denoting $q = a^2$, we see that $\dot{q}$ is a linear function of $\tau$ and hence $q$ is a quadratic function $\tau$. Imposing the initial conditions $q(\tau_0) = q_0$ and $\dot{q}(\tau_0)=\dot{q}_0$, the solution is given by
	\ba
q(\tau) &=& - \qty(\tau-\tau_0)^2 +  \dot{q}_0\qty(\tau-\tau_0) +  q_0\,.
	\ea
For the evaluation of the path integral we have to translate the initial value problem $a(0)=a_0$ and $\dot{a}(0)=a_0$ into the boundary value problem $a(\tau_0)=a_0$ and $a(\tau_1)=a_1$. This can be done by tuning the initial expansion rate of the squared throat radius as
\ba\label{eq:aDot0Sol}
\dot{q}_0 &=& \frac{q_0 - q_1 - \qty( \Delta\tau)^2 }{\Delta \tau}\,,
\ea
where we remind that $\Delta\tau = \tau_1 - \tau_0$,
	such that
	\ba\label{eq:aSol}
		q_\sigma(\tau)^2 &=& - \qty(\tau - \tau_0)^2 +  \qty[\frac{q_0 - q_1 }{\Delta \tau}-   \Delta\tau](\tau - \tau_0) + q_0\,.
	\ea
	  The right hand side describes a parabola and is non-negative for $\tau \in [\tau_0,\tau_1]$ and $ a_0 < a_1$, which we will assume. As the wormhole geometry is only defined for non-negative throat radii, we consider only the positive solution $a_\sigma(\tau) = + \sqrt{q_\sigma(\tau)}$ and  from now on omit the subscripts $\sigma$ labeling a given saddle.

The previous expressions allow us to obtain the wormhole  onshell effective action $S$ 
	as a function of $a_0$, $a_1$ and $\Delta \tau$. To that end, it is useful to split~\eqref{eq:ActionEvaluated} as follows,
	\ba
	S &=& S_1 + S_2\,,
	\ea
where
\ba
	S_1 &=& 2 \int_{\tau_0}^{\tau_1} \dd{\tau}  a \dot{a}\,\text{arcsinh}(\dot{a})\,,\\
	S_2 &=& - 2  \int_{\tau_0}^{\tau_1} \dd{\tau} a \sqrt{1 + \dot{a}^2}\,.
\ea
The first term can be integrated by parts to yield
\ba
S_1 &=& \eval{a^2 \,\text{arcsinh}\qty(\dot{a})}_{\tau_0}^{\tau_1} - \int_{\tau_0}^{\tau_1} \dd{\tau} \frac{a^2 \,\ddot{a}}{\sqrt{1+\dot{a}^2}}\,.
\ea
Now we use that onshell $1+ \dot{a}^2 = - a \ddot{a}$, cf.~equation~\eqref{eq:ELEOM}, and therefore
\ba
 - \int_{\tau_0}^{\tau_1} \dd{\tau} \frac{a^2 \,\ddot{a}}{\sqrt{1+\dot{a}^2}} &=&   \int_{\tau_0}^{\tau_1} \dd{\tau} a \sqrt{1+\dot{a}^2}\,.
\ea
Thus $S_1$ and $S_2$ sum to
\ba
S &=& \eval{a^2 \,\text{arcsinh}\qty(\dot{a})}_{\tau_0}^{\tau_1}  -  \int_{\tau_0}^{\tau_1} \dd{\tau} a \sqrt{1+\dot{a}^2}\,,
\ea
whereby for the solution specified in~\eqref{eq:aSol} the remaining integrand evaluates to a constant,
\ba
a \sqrt{1+\dot{a}^2} &=& \sqrt{q + \frac{1}{4}\dot{q}^2}\,\,=\,\, \sqrt{q_0 + \frac{1}{4}\dot{q}_0^2}\,, 
\ea
with $\dot{q}_0(q_0,q_1,\Delta \tau)$ given in~\eqref{eq:aDot0Sol}.
Altogether, in general the action evaluates to
\ba\label{eq:SOnshellGeneric}
S\qty(a_0,a_1,\Delta \tau)&=&   \eval{q \,\text{arcsinh}\qty(\frac{\dot{q}}{2 \sqrt{q}})}_{\tau_0}^{\tau_1}  -  \sqrt{q_0 + \frac{1}{4}\dot{q}_0^2}\,\Delta \tau \,.
\ea
For example, setting $\Delta \tau =1$ explicitly leads to
\ba\label{eq:SOnshell}
S\qty(a_0,a_1,1) &=& -a_0^2\,\text{arcsinh}\qty(\frac{1 + \qty(a_1^2 -a_0^2)}{2 a_0})  - a_1^2\,\text{arcsinh}\qty(\frac{1 - \qty(a_1^2 -a_0^2)}{2 a_1}) \nn\\
&-& \frac{1}{2}\sqrt{\qty(a_0^2 - a_1^2)^2 + 2 \qty(a_0^2 + a_1^2) + 1}\,.
\ea
Figure~\ref{Fig:SJ} (left plot) shows the onshell action $S$ for a choice of proper-time interval $\Delta \tau = 1$. Note that the action is real.
\\
	
In the final step, the corresponding Hessian factor $\mathcal{J}$ defined in~\eqref{eq:PathIntegralApproximation} can be obtained explicitly by taking the derivative of~\eqref{eq:SOnshellGeneric} with respect to $a_1$ and $a_0$, which results in
\ba
\pdv[2]{S\qty(a_0,a_1,\Delta \tau)}{a_0}{a_1}&=& \frac{- 2 a_0 a_1}{\sqrt{(a_0 + a_1)^2 + \Delta \tau^2} \sqrt{(a_0 - a_1)^2 + \Delta \tau^2}}\,.
\ea
	Figure~\ref{Fig:SJ} (right plot) shows the real and imaginary parts of $\mathcal{J}$ for $\Delta \tau = 1$. It should be observed that
	\ba
	\lim_{a_0 \to 0} \mathcal{J} = 0
	\ea
	and similarly in the limit $a_1 \to 0$. Thus the Hessian determinant suppresses topology-changing processes in which a wormhole is created or destroyed. More precisely, the fact that $\mathcal{J}$ becomes zero for these types of transitions signalizes that the WKP approximation fails. By analogy with a non-relativistic particle, this situation corresponds to placing an infinitely steep potential wall at $a =0$, leading to the boundary condition $\psi(0) =0$ on the wave function. The same type of behaviour has been observed in a canonical quantization by solving the Wheeler-de-Witt type equation $\hat{H}\psi(a) = 0$ in~\cite{Visser:1989am,Visser:1990wi}.

	We can combine $S$ and $\mathcal{J}$ to evaluate the propagator according to~\eqref{eq:PathIntegralApproximation}. The result is shown in Figure~\ref{Fig:G} for $\Delta \tau = 1$. In addition the probability distribution $\abs{G}^2 = G G^*$ is shown in Figure~\ref{Fig:GGStar}. As already anticipated, we observe that the probability of topology change goes to zero.

\section{Thermodynamics in the presence of a matter thin shell}	\label{Sec:Thermodynamics}

Our goal in the following is to derive a gravitational thermodynamics due the  existence of a wormhole throat in spacetime. To that end we will make use of the path-integral representation of the canonical partition function. The throat $\Sigma$ represents a timelike junction surface in the $4$-dimensional background geometry. We start by Wick rotating the action~\eqref{eq:ActionEvaluated} via the replacement $\tau \to -\imath \tau_{\text{E}}$, where $\tau_{\text{E}}$ represents Euclidean proper time. This leads to the Euclidean effective action 
\ba\label{eq:ActionEvaluatedEuclidean}
\imath S_{\text{E}}[a(\tau_\text{E})] = \imath \int  \dd{\tau}_{\text{E}}  \qty( 2  a \,a'\,\arcsin({ a'}) + 2 a \,\sqrt{1-a'^2} ) \,,
\ea
where a prime denotes the derivative with respect to $\tau_\text{E}$. The thermodynamic gravitational canonical partition function is formally defined as an integral
\ba\label{eq:ZBeta}
Z(\beta) = \int_{a(\tau_{\text{E}}) \,= \,a(\tau_{\text{E}} + \beta )}\mathcal{D}a \, e^{- S_{\text{E}}[a(\tau_{\text{E}})]}
\ea
over periodic functions $a(\tau_\text{E})$ with period $\beta$. From the canonical partition function the free energy $F$ can be computed as follows,
\ba
\beta F = S_{\text{E}} = -\ln(Z)\,.
\ea
In the following we will assume that~\eqref{eq:ZBeta} is dominated by classical saddles. The Euler-Lagrange equations arising by varying the Euclidean action~\eqref{eq:ActionEvaluatedEuclidean} are given by 
\ba\label{eq:ODEa}
a a'' + a'^2 - 1 = 0\,.
\ea
This differential equation can be integrated twice to obtain solutions in the form
\ba
a(\tau_\text{E})^2 = \tau_{\text{E}}^2 + 2 a_0 a'_0 \tau_{\text{E}} + a_0^2\,.
\ea
These solutions are not periodic in $\tau_\text{E}$, or in other words their period is $\beta \to \infty$. Therefrom one would associate a zero temperature $T = \frac{1}{\beta} \to 0$.  \\

The situation changes if we modify the wormhole geometry by adding matter in the form of a thin shell confined to lie on the throat of the wormhole, see e.g.~\cite{Visser:1989kg,Visser:1990wi,Visser:1995cc}. The full spacetime energy-momentum tensor can be written in the form
\ba\label{eq:TTensor}
T\indices{^\mu_\nu}(x)&=&  {T_{+}}\indices{^\mu_\nu}\Theta(\eta) + {T_{-}}\indices{^\mu_\nu}\Theta(-\eta) + S\indices{^\mu_\nu}\delta(\eta)\,,
\ea
where the energy-momentum tensors ${T_{\pm}}\indices{^\mu_\nu}$ on both sides of the throat  vanish identically and $S\indices{^\mu_\nu}$ denotes the surface stress-energy tensor. Combining~\eqref{eq:TTensor} with the expressions of the Ricci tensor~\eqref{eq:RicciTensor} and Ricci scalar~\eqref{eq:RicciScalar}, the Einstein field equations $G_{\mu\nu} = 8 \pi T_{\mu\nu}$ projected onto the throat imply 
\ba\label{eq:FieldEquations}
S\indices{^i_j} = - \frac{1}{8 \pi} \qty[\Delta {K_{\Sigma}}\indices{^i_j} - \Tr(\Delta K_\Sigma)\delta\indices{^i_j}]\,.
\ea
The surface stress-energy tensor for a thin shell with surface energy density $\sigma $ and surface tension $\zeta = -p$, corresponding to the negative surface pressure $p$,  is given by
\ba\label{eq:S}
S\indices{^i_j} = \text{diag}\qty{S\indices{^\tau_\tau}, S\indices{^\theta_\theta}, S\indices{^\theta_\theta}}= \text{diag}\qty{-\sigma,-\zeta,-\zeta}\,.
\ea
The field equations~\eqref{eq:FieldEquations} can be recast into the pair 
\ba
\sigma & =& - \frac{1}{2 \pi}{ K_{\Sigma}}\indices{^\theta_\theta} \,= \,- \frac{1 }{2 \pi a}\sqrt{1 + \dot{a}^2} \,,
\label{eq:Einstein1}\\
\zeta &=& -\frac{1}{4 \pi}\qty({K_\Sigma}\indices{^\tau_\tau} + { K_\Sigma}\indices{^\theta_\theta} ) \,= \, -\frac{1}{4 \pi a  \sqrt{1+\dot{a}^2}}\qty( a \ddot{a} + \dot{a}^2 +1) 
\label{eq:Einstein2}\,,
\ea
where we have made use of~\eqref{eq:ExtrinsicCurvatureComponents1}--\eqref{eq:ExtrinsicCurvatureComponents2} and~\eqref{eq:TraceDiscontinuityExtrinsicCurvature}. Instead of~\eqref{eq:Einstein1}--\eqref{eq:Einstein2}, one may equivalently consider the equations
\ba
4 \pi^2 \sigma^2 a^2 - \dot{a}^2 - 1  & =& 0\,,
\label{eq:Einstein1New}\\
\dot{\sigma} +2 \qty(\sigma - \zeta)\frac{\dot{a}}{a} &=& 0\,.
\label{eq:Einstein2New}
\ea
The last equation implies the conservation equation
\ba\label{eq:ConservationEq}
\derivative{}{\tau}\qty(4 \pi  \sigma a^2) = 4 \pi \zeta \,\derivative{}{\tau}\qty(a^2)\,.
\ea

In order to perform a path-integral analysis, the main question is whether and how the above equations can be obtained from a variational action principle. This question is easy to answer in the absence of surface tension. To that end we note that the conservation equation~\eqref{eq:ConservationEq}, when $\zeta=0$, implies that the effective mass of the dust shell
\ba\label{eq:MuMass}
\mu = 4 \pi \sigma a^2
\ea 
is a constant of motion. In particular,~\eqref{eq:Einstein1} and~\eqref{eq:Einstein2} in this case reduce to
\ba
2 a \sqrt{1+\dot{a}^2} + \mu  &=& 0\,,\label{eq:Einstein1NoP}\\
a \ddot{a} + \dot{a}^2 + 1 &=& 0\label{eq:Einstein2NoP}\,.
\ea
We recognize the last equation as the Euler-Lagrange equation~\eqref{eq:ELEOM} derived from the Lorentzian wormhole action~\eqref{eq:ActionEvaluated}, whereas the first equation can be written as
\ba\label{eq:HPrime}
H'(a,\dot{a}) \equiv  H(a,\dot{a}) + \mu = 0\,,
\ea
with the canonical Hamiltonian $H(a,\dot{a})$ given in~\eqref{eq:Hamiltonian}. This relation should not to be viewed as a constraint, but rather as an integrated equation of motion whose derivative with respect to $\tau$ returns back the Euler-Lagrange equation of motion~\eqref{eq:Einstein2NoP}. It is easy to see that $H'$ coincides with the canonical Hamiltonian derived from the action~\eqref{eq:ActionEvaluated}, supplemented by a matter term in the form
\ba\label{eq:ActionEvaluatedInclMatter}
S &=& S_{\text{wormhole}} - \int \dd{\tau}\mu\,.
\ea
It should be emphasized that this action is not reparametrization invariant and the Israel-Lanczos equation $H' = 0$ in~\eqref{eq:HPrime} does not follow by variation of this action. The observation that differentiating $H' = 0$ with respect to $\tau$ leads to the Euler-Lagrange equations applies similarly to the non-relativstic particle, cf.~Appendix~\ref{App:NonRelativsticParticle}. In particular, it is clear that adding matter with non-zero surface energy density but zero surface tension, to motivate the action~\eqref{eq:ActionEvaluatedInclMatter}, does not modify the Euler-Lagrange equations, and therefore does not solve the above problem of non-periodicity of solutions to the Euclidean theory.\\

In the following we will proceed by assuming that there exists an action  of the form
\ba
S = S_{\text{wormhole}} + S_{\text{matter}} =  \int \dd{\tau} \qty(2 a \dot{a}\,\text{arcsinh}(\dot{a}) -2 a \sqrt{1+\dot{a}^2}) + S_{\text{matter}}\,,
\ea
which gives rise to the~\eqref{eq:Einstein1New}--\eqref{eq:Einstein2New} as the dynamical equations of motion for a priori generic surface energy density $\sigma$ and surface tension $\zeta$. Prescriptions on how such an action can be generated may be found in~\cite{Brown:1992kc,Hayward:1990tz}. With this assumption we consider the Wick rotation  of the above action as a result of the replacement $\tau \to -\imath \tau_{\text{E}}$, 
\ba\label{eq:SEFull}
\imath S_{\text{E}} = \imath S_{\text{E wormhole}} + \imath S_{\text{E matter}} =  \imath \int \dd{\tau_{\text{E}}} \qty(2 a a'\,\arcsin(a') + 2 a \sqrt{1-a'^2}) + S_{\text{E matter}}\,.
\ea
Therefrom we define the thermodynamic gravitational partition function $Z(\beta)$ analogously as in~\eqref{eq:ZBeta}, where $S$ denotes the effective action onshell of matter degrees of freedom. This functional integral is then subjected to a saddle-point approximation
such that, in effect, we are interested in solutions to the Euclideanized field equations which arise from~\eqref{eq:Einstein1}--\eqref{eq:Einstein2} by the replacement $\dot{a}^2 \to - a'^2$ and $\ddot{a} \to -a''$,
\ba
\sigma &=&  -\frac{1}{2 \pi a}\sqrt{1-a'^2}\,,\label{eq:Einstein1Euclidean}\\
\zeta &=&  \frac{1}{4 \pi a \sqrt{1-a'^2}}\qty(a a'' + a'^2 - 1)\label{eq:Einstein2Euclidean}\,.
\ea
Concretely, our primary question is whether these are periodic with finite period $\beta$, such that in the framework of a Euclidean path integral over periodic functions a gravitational temperature $T = \frac{1}{\beta}$ can be defined. It should be emphasized that this temperature and  the thereby defined gravitational thermodynamics of spacetime are distinct from the  standard thermodynamics of the matter shell.\\

Instead of equations~\eqref{eq:Einstein1Euclidean}--\eqref{eq:Einstein2Euclidean}, similarly as before, we consider the two equations
\ba
4 \pi^2 \sigma^2 a^2 + a'^2 - 1 &=& 0\,, \label{eq:Einstein1EuclideanNew}\\
\sigma' + 2 \qty(\sigma - \zeta)   \frac{a'}{a} &=& 0\,.\label{eq:Einstein2EuclideanNew}
\ea
These can be recast in a form that is reminiscent of  the Euclidean first Friedmann equation and the continuity equation in a $2+1$ dimensional FLRW spacetime with spatial curvature $k=1$,
\ba
\qty(\frac{a'}{a})^2  + 4 \pi^2 \sigma^2   - \frac{1}{a^2} &=& 0\,, \label{eq:Einstein1EuclideanNewFLRW}\\
\sigma' + 2 \qty(\sigma +p)   \frac{a'}{a} &=& 0\,.\label{eq:Einstein2EuclideanNewFLRW}
\ea
However, different from the first Friedmann equation, the above first equation involves the square of the energy density $\sigma^2$ rather than $\sigma$.
In the following we will impose  the barotropic equation of state
\ba \label{eq:EOS}
p(\sigma) = \omega \sigma
\ea
with a non-zero constant equation of state parameter $\omega $. Equation~\eqref{eq:Einstein2EuclideanNewFLRW}   can be separated and solved in the form
\ba
\frac{\dd{\sigma}}{\sigma} =- 2(1+\omega) \frac{\dd{a}}{a} \,\,\,\quad \Rightarrow \quad \,\,\, \ln(\frac{\sigma}{\sigma_0})= -2(1+\omega) \ln(a) \,,
\ea
where $\sigma_0 < 0$ is an integration constant.
Thus the energy density $\sigma$ as a function of the throat radius $a$ is given by
\ba\label{eq:Sigmaa}
\sigma(a) = \sigma_0 a^{-2 (1+\omega)}\,.
\ea
Inserting the previous expression into equation~\eqref{eq:Einstein1EuclideanNewFLRW}, we are left with
\ba\label{eq:ODEEuclideanReduced}
a'^2 + \gamma(\sigma_0) a^{\delta(\omega)}= 1 \,\,\,\quad \Leftrightarrow \quad \,\,\, T + V = E\,,
\ea
where
\ba 
\gamma(\sigma_0) = 4 \pi^2 \sigma_0^2\,\, > \,\,0 \,\,\,\quad \text{and} \,\,\,\quad
\delta(\omega) = -2(1+2\omega)\,.
\ea
We will henceforth assume $\omega < -\frac{1}{2}$. The relation~\eqref{eq:ODEEuclideanReduced} can be interpreted as a standard classical mechanical equation reflecting the conservation of energy $E$ as the sum of kinetic energy $T$ and potential energy $V$. From~\eqref{eq:ODEEuclideanReduced} we obtain 
\ba\label{eq:aPrime}
a' = \pm \sqrt{1 - \gamma a^\delta}\,.
\ea
Bounded motion will have a Euclidean proper-time period given by
\ba
\beta &\equiv & \Delta \tau_{\text{E}} =
 -2\int_{-a_{\text{max}}}^{0} \frac{\dd{a} }{\sqrt{1- \gamma a^{\delta}}} + 2 \int_0^{+a_\text{max}} \frac{\dd{a} }{\sqrt{1- \gamma a^{\delta}}} \nn\\
&=& \gamma^{-\frac{1}{\delta}}  \qty[-2\int_{-1}^{0} \frac{ \dd{x} }{\sqrt{1-  x^{\delta}}} + 2 \int_{0}^{+1} \frac{ \dd{x} }{\sqrt{1-  x^{\delta}}}]\nn\\
& \equiv &   \gamma^{-\frac{1}{\delta}}\,\, \mathcal{I}_1(\delta)\,, \label{eq:DeltaTauE}
\ea
where $\pm a_{\text{max}} = \pm \gamma^{-\frac{1}{\delta}}$ denote the classical turning points at which $a' =0$, and in the second line we have performed the change of variable $a \to \gamma^{-\frac{1}{\delta}}x$. The above integrals can be solved in terms of hypergeometric and Gamma functions. The Euclidean proper-time period $\beta$ implies a temperature whose dependence on the zero-point surface energy density  $\sigma_0$ is
\ba\label{eq:TSigma0}
T \,\,=\,\, \frac{1}{\beta} \,\,\propto\,\, \gamma^{\frac{1}{\delta}}\,\, \sim \,\, \abs{\sigma_0}^{-\frac{1}{ 1+2\omega}}\,.
\ea
It should be observed that this gravitational temperature is constant and in particular distinct from a standard matter temperature satisfying the Stefan-Boltzmann law $\sigma \propto T_{\text{matter}}^4$ in time. In fact, the above temperature can be understood as a purely geometric quantity due to the discontiuity of the extrinsic curvature across the junction,
\ba
\abs{\Delta {K_{\Sigma}}\indices{^\theta_\theta}} = \abs{\frac{1}{a}\sqrt{1-a'^2}} = \abs{\sqrt{\gamma}\, a^{\frac{\delta}{2}-1}} = 2 \pi \abs{\sigma_0} \,a^{-2 (1+\omega)}\,,
\ea
which implies
\ba\label{eq:TDeltaK}
T \,\,\propto \,\,\abs{ \Delta {K_{\Sigma}}\indices{^\theta_\theta} \,a^{2(1+\omega) } }^{-\frac{1}{1+2\omega}} \,.
\ea
The previous considerations allow us to evaluate the gravitational contribution to the Euclidean action~\eqref{eq:SEFull} onshell. Inserting~\eqref{eq:aPrime} into~\eqref{eq:SEFull} results in
\ba
 S_{\text{E wormhole}} &=& \int_0^{\beta} \dd{\tau_{\text{E}}}\qty(2 a \sqrt{1-\gamma a^{\delta}} \,\arcsin(\sqrt{1-\gamma a^{\delta}}) + 2 a \sqrt{\gamma} a^{\frac{\delta}{2}}) \nn\\
&=& -2\int_{-a_{\text{max}}}^{0} \dd{a} 2 a \qty( \arcsin(\sqrt{1-\gamma a^\delta}) +  \frac{ \sqrt{\gamma} a^{\frac{\delta}{2}}}{\sqrt{1-\gamma a^{\delta}}} )\nn\\
&{}& + 2\int_{0}^{+a_{\text{max}}} \dd{a} 2 a \qty( \arcsin(\sqrt{1-\gamma a^\delta}) +  \frac{ \sqrt{\gamma} a^{\frac{\delta}{2}}}{\sqrt{1-\gamma a^{\delta}}} )\nn\\
&=& \gamma^{-\frac{2}{\delta}}\Bigg[ -2\int_{-1}^0 \dd{x} 2x  \qty( \arcsin(\sqrt{1-x^{\delta}}) + \frac{   x^{\frac{\delta}{2}}}{\sqrt{1-x^\delta}})\,\,\,\nn\\
&{}&  \quad \quad \,\,\,+2\int_{0}^{+1} \dd{x} 2x  \qty( \arcsin(\sqrt{1-x^{\delta}}) + \frac{   x^{\frac{\delta}{2}}}{\sqrt{1-x^\delta}})\Bigg]\nn\\
&\equiv& \gamma^{-\frac{2}{\delta}}\, \mathcal{I}_2(\delta)\,\, \propto \,\,\beta^2\,.
\ea
Combining this result with~\eqref{eq:DeltaTauE} implies a gravitational contribution to the free energy
\ba\label{eq:FreeEnergy}
F \,\,\propto \,\,\beta = \gamma^{-\frac{1}{\delta}} \,\,\propto \,\,\abs{\sigma_0}^{\frac{1}{ 1+2\omega}}\,.
\ea
Therefrom we obtain the gravitational entropy according to the canonical ensemble
\ba\label{eq:Entropy}
S \,\,= \,\,(1-\beta \partial_\beta)\ln(Z) = - (1-\beta \partial_\beta)\qty[\beta F] \,\,\propto \,\,\beta^2 \,\,\propto\,\, \abs{\sigma_0}^{\frac{2}{1+2\omega}} \,.
\ea
Altogether, we conclude  that the temperature and entropy associated with the gravitational degrees of freedom are constants depending on the zero-point surface energy density $\sigma_0$ which parametrizes the jump of the extrinsic curvature, through a power law specific to the equation of state parameter $\omega$ defined in~\eqref{eq:EOS}. 

We can use the previous results to write a thermodynamic first law by making use of the effective mass of the dust shell defined in~\eqref{eq:MuMass},
\ba
\mu(a) = 4 \pi \sigma(a)a^2 = \sigma(a) A\,,
\ea
where $A = 4 \pi a^2$ is the cross-sectional area of the throat at radius $a$. With $\dd{S}=0$ it follows that
\ba\label{eq:FirstLaw}
\dd{\mu} = T \dd{S} - p \dd{V} \,\,\,\quad \Leftrightarrow \quad \,\,\, \qty[\pdv{\sigma}{A}A+\sigma] \dd{A} = - p\dd{V} \,,
\ea
where the volume $V$ of the system is the $2$-dimensional area $A$. For this identity to hold, it must be
\ba
p = \omega \sigma = -\qty[\pdv{\sigma}{A}A+\sigma] 
 \,\,\, \quad \Rightarrow \,\,\, \quad - \frac{1}{1+\omega} \frac{\dd{\sigma}}{\sigma} = \frac{\dd{A}}{A}\,.
\ea
Integrating on both sides leads to
\ba\label{eq:SigmaA}
\sigma(A) \,\,\propto \,\, A^{-(1+\omega)}\,,
\ea
which is equivalent to~\eqref{eq:Sigmaa}. This confirms the validity of the first law~\eqref{eq:FirstLaw}, which can be understood as a differential version of the conservation equation~\eqref{eq:ConservationEq}. In the following we consider two special cases.\\

\noindent {\it Equation of state $p(\sigma) =  - \sigma$}\\

\noindent Physically this case can be interpreted as a classical membrane. From~\eqref{eq:Einstein1EuclideanNew} it follows that the surface energy density $\sigma = \sigma_0$ is a constant. Then the first-order differential equation~\eqref{eq:Einstein1EuclideanNew} can be integrated to obtain
\ba
a(\tau_\text{E}) = \pm \frac{1}{2 \pi \sigma_0}\sin(2 \pi \sigma_0(\tau_{\text{E}} - c))\,,
\ea
where $c$ is an integration constant. These solutions are periodic in $\tau_{\text{E}}$ with period
\ba
\beta \,\,\equiv \,\,\Delta \tau_{\text{E}} =  \frac{1}{\abs{ \sigma_0}}\,.
\ea
Therefrom one may associate a temperature given by
\ba
T \,\,= \,\,\frac{1}{\beta} = \abs{\sigma_0} \,\,\propto \,\,\abs{ \Delta {K_{\Sigma}}\indices{^\theta_\theta}}\,.
\ea
This result is in agreement with~\eqref{eq:TSigma0}--\eqref{eq:TDeltaK} for $\omega = -1$. The free energy~\eqref{eq:FreeEnergy} and entropy~\eqref{eq:Entropy} are given by
\ba
F \,\,\propto \,\, \abs{\sigma_0}^{-1} \,\,\propto \,\, \abs{ \Delta {K_{\Sigma}}\indices{^\theta_\theta}}^{-1}\,\,\,\quad \text{and} \quad \,\,\, S \,\,\propto\,\, \abs{\sigma_0}^{-2}  \,\,\propto \,\, \abs{ \Delta {K_{\Sigma}}\indices{^\theta_\theta}}^{-2} \,.
\ea
The thermodynamic first law~\eqref{eq:FirstLaw} becomes
\ba
\dd{\mu} = T \dd{S} - p \dd{V} \,\,\,\quad \Leftrightarrow \quad \,\,\, \sigma_0 \dd{A} = - p\dd{A}\,. 
\ea
\\

\noindent {\it Non-zero energy density $\sigma$ and zero pressure $p$}\\

\noindent This case can be interpreted as pressureless dust confined on the throat. The Euclideanized field equations~\eqref{eq:Einstein1EuclideanNew}--\eqref{eq:Einstein2EuclideanNew} can be solved in the form
\ba
a(\tau_\text{E})^2 &=& \tau_{\text{E}}^2 \pm \sqrt{\qty(1 - 4 \pi^2 \sigma_0^2 a_0^2) a_0^2} \, \tau_{\text{E}} + a_0^2\,,\\
\sigma(\tau_{\text{E}}) &=& \sigma_0 \frac{a_0^2}{a(\tau_{\text{E}}^2)}\,.
\ea
These solutions are not periodic in $\tau_{\text{E}}$, as previously anticipated. Therefrom  we conclude that a matter shell with vanishing surface tension and non-zero energy density alone is not sufficient to define a finite temperature from the perspective of a Euclidean path integral over periodic functions.

\section{Discussion}\label{Sec:Discussion}

Wormholes are of major relevance for finding answers to key open questions in quantum gravity, such as the question of topology change, the role of global symmetries, and the microscopic counting of states that leads to black-hole entropy. Wormhole spacetimes, 
by definition, must violate the null convergence condition in order to ensure the defocusing of light rays impinging on the throat. If the throat satisfies dynamical equations, it is meaningful ask about the limit of vanishing throat radius, corresponding physically to the creation of a wormhole which connects two initially separated universes, or to the complete evaporation of an existing wormhole when two connected universes merge into one. Both types of processes represent topology change. Expecting such extreme behavior to be dictated by quantum theory, one is lead to consider the quantum dynamics of the wormhole throat. In practice this amounts to asking:  {\it What is the probability of transition between a quantum state with zero throat radius and another one with finite throat radius?}

We have addressed this question for a cut-and-paste Lorentzian wormhole obtained by gluing two identically cut Minkowski spacetimes along a timelike junction surface. Concretely, we have focused on a path-integral representation for the propagator associated with the above transition, with an effective action obtained by evaluating the Einstein action for the wormhole spacetime taking into account the dynamics of the junction surface. Despite such a path integral resembling closely a quantum-gravitational path integral, in which the Einstein  action is ought to represent just the leading-order contribution in a series of higher-curvature terms, the reduced minisuperspace wormhole action is a complicated non-polynomial function of derivatives due  to the delta-function singularity of the Riemann curvature on the throat. Moreover, as the wormhole construction features generically background geometries on each side of the throat, making the meaning of a global $3+1$ split ambiguous, we do not view such path integral as being in any obvious way related to a path integral for general relativity. Instead, we have quantized the effective action governing the dynamics of the throat as a non-relativstic one-degree-of-freedom  classical mechanical  system on its own.

Based on a saddle-point approximation dominated by a  single classical parabolic trajectory for the squared throat radius, we have shown that topology change is efficiently suppressed by the Hessian determinant. This conclusion aligns with results of the works~\cite{Visser:1989am,Visser:1990wh,Visser:1990wi} which have considered a canonical quantization by solving the Wheeler-de-Witt type equation $\hat{H}\psi = 0$. As previously emphasized, however, the constraint $H=0$ does not arise from the effective action for the wormhole but instead represents one of the Israel-Lanczos dynamical equations on the throat. Understanding if and how this equation reflects reparametrization invariance of general relativity is a key step towards establishing a possible relation of the non-polynomial effective action for the throat radius to a minisuperspace $3+1$ path integral for general relativity. In particular, it remains to clarify the degrees of freedom over which such a path integral should sum over. This knowledge may in turn allow for attributing a statistical mechanical entropy to the throat surface, defined as a Lorentzian path integral over closed paths on the reduced phase space~\cite{,Banihashemi:2024aal,Banihashemi:2024weu,Dittrich:2024awu}. 

In the current lack of understanding about the above questions, we have proceeded heuristically in associating a gravitational thermodynamics  to the wormhole throat in spacetime. To that end, we have considered the Wick rotation of the Israel-Lanczos equations in the presence of a thin-shell source, which are assumed to originate as the dynamical equations of motion from a Euclidean effective gravity-matter action. The thereby obtained gravitational temperature and entropy are constants sourced by the discontinuity of the extrinsic curvature across the wormhole throat, with an $S \sim T^{-2}$ dependence characteristic to gravitational systems with horizons. It is appealing  to investigate the gravitational thermodynamic properties of junction surfaces further and see how they may reveal essential features of thermodynamics for generic gravitational systems, including  black holes and de Sitter space.

\begin{acknowledgments}
We thank Job Feldbrugge and Raymond Isichei for discussions. Both authors acknowledge support by STFC Consolidated Grant ST/T000791/1.

\end{acknowledgments}

\appendix

\section{Reparametrization invariance on the extended phase space}\label{App:ReparametrizationInvariance}

The effective action~\eqref{eq:ActionEvaluated} is not reparametrization invariant. Reparametrization invariance can be regained on the extended configuration space obtained by promoting the coordinate $\tau$ to a new configuration variable $\tau \to \tau(t)$, in addition to $a(t)$. Here $t$ is an apriori arbitrary evolution parameter which can be related e.g.~to the background Minkowski time. Thereby we obtain
\ba\label{eq:ActionExtended}
S_{\text{extended}}[a(t),\tau(t)] &=& \int \dd{t} L_{\text{extended}}\qty(a,a',\tau')= \int \dd{t} \tau' L\qty(a, \frac{a'}{\tau'}) \nn\\
&=& \int \dd{t} \qty(2 a a' \sinh^{-1}\qty({\frac{a'}{\tau'}}) - 2 a \tau' \sqrt{1 + \qty(\frac{a'}{\tau'})^2} )\,,
\ea
where a prime denotes the derivative with respect to $t$. The extended Lagrangian is homegenous of degree one in the velocities $(\tau',a')$ and therefore reparametrization invariant. The conjugate momenta are given by
\ba
p_\tau &=& \frac{\partial L_{\text{extended}}}{\partial \tau'} = L\qty(a, \frac{a'}{\tau'}) - \frac{a'}{\tau'} \pdv{L\qty(a, \frac{a'}{\tau'})}{\qty(\frac{a'}{\tau'})} = L\qty(a,\dot{a})- p \dot{a} = -H(a,p)\,,\\
p_a &=&  \frac{\partial L_{\text{extended}}}{\partial a'} = \pdv{L\qty(a, \frac{a'}{\tau'})}{\qty(\frac{a'}{\tau'})} =  \pdv{L\qty(a, \dot{a})}{\dot{a}}  = p\,.
\ea
Thus $p_\tau$, $p_a$ and $a$ have to satisfy the (Hamiltonian) constraint
\be
\mathcal{H} = p_\tau + H(a,p) \approx 0\,.
\ee
The extended Hamiltonian on the extended phase space is defined as
\be
H_{\text{extended}} = p_a a' + p_\tau \tau' - L_{\text{extended}} = \tau' \mathcal{H}\,.
\ee
Herewith the extended action~\eqref{eq:ActionExtended} can be written as
\be
S_{\text{extended}}[a(t),\tau(t)] = \int \dd{t} \qty(p_a a' + p_\tau \tau' - N(t) \mathcal{H})\,\,\,\quad \text{where} \quad N(t) = \derivative{ \tau(t)}{t}
\ee
is the lapse function, which measures how much proper time $\tau$ has elapsed with respect to the evolution parameter $t$. Fixing the gauge to $\tau = t$ we recover the action~\eqref{eq:ActionEvaluated}.

\section{Non-relativistic particle propagator}\label{App:NonRelativsticParticle}

 In the following we review the derivation of the non-relativistic particle propagator expressed as a path integral. A non-relativistic particle in a quadratic potential is described by the action
 \ba\label{eq:ActionNonRelParticle}
 S[x(\tau)] = \frac{1}{2}\int \dd{\tau}\qty(\dot{x}^2 -x^2)\,.
 \ea
 Varying with respect to $x(\tau)$ leads to the Euler-Lagrange equations
\ba\label{eq:EOMELParticle}
  \derivative{}{\tau}\qty(\pdv{L}{\dot{x}}) - \pdv{L}{x} \,=\,0  \,\,\, \quad \Rightarrow \quad \,\,\, \ddot{x} + x\,= 0\,.
\ea
The momentum conjugated to $x$ in~\eqref{eq:ActionNonRelParticle} is
\ba\label{eq:P}
p = \pdv{L}{\dot{x}} = \dot{x}\,.
\ea
Therefore the canonical Hamiltonian is given by
\ba
H(x,p) = p \dot{x} - L = \frac{1}{2}\qty(p^2 + x^2) \,\,\,\quad \Leftrightarrow \quad \,\,\, H(x,\dot{x}) = \frac{1}{2}\qty(\dot{x}^2 + x^2)\,.
\ea
It should be observed that differentiating the equation $H=0$ with respect to $\tau$ returns the Euler-Lagrange equations. The analogous observation applies to the wormhole Hamiltonian~\eqref{eq:Hamiltonian}.\\

\begin{figure}[t]
	\centering
	\includegraphics[width=.47\textwidth]{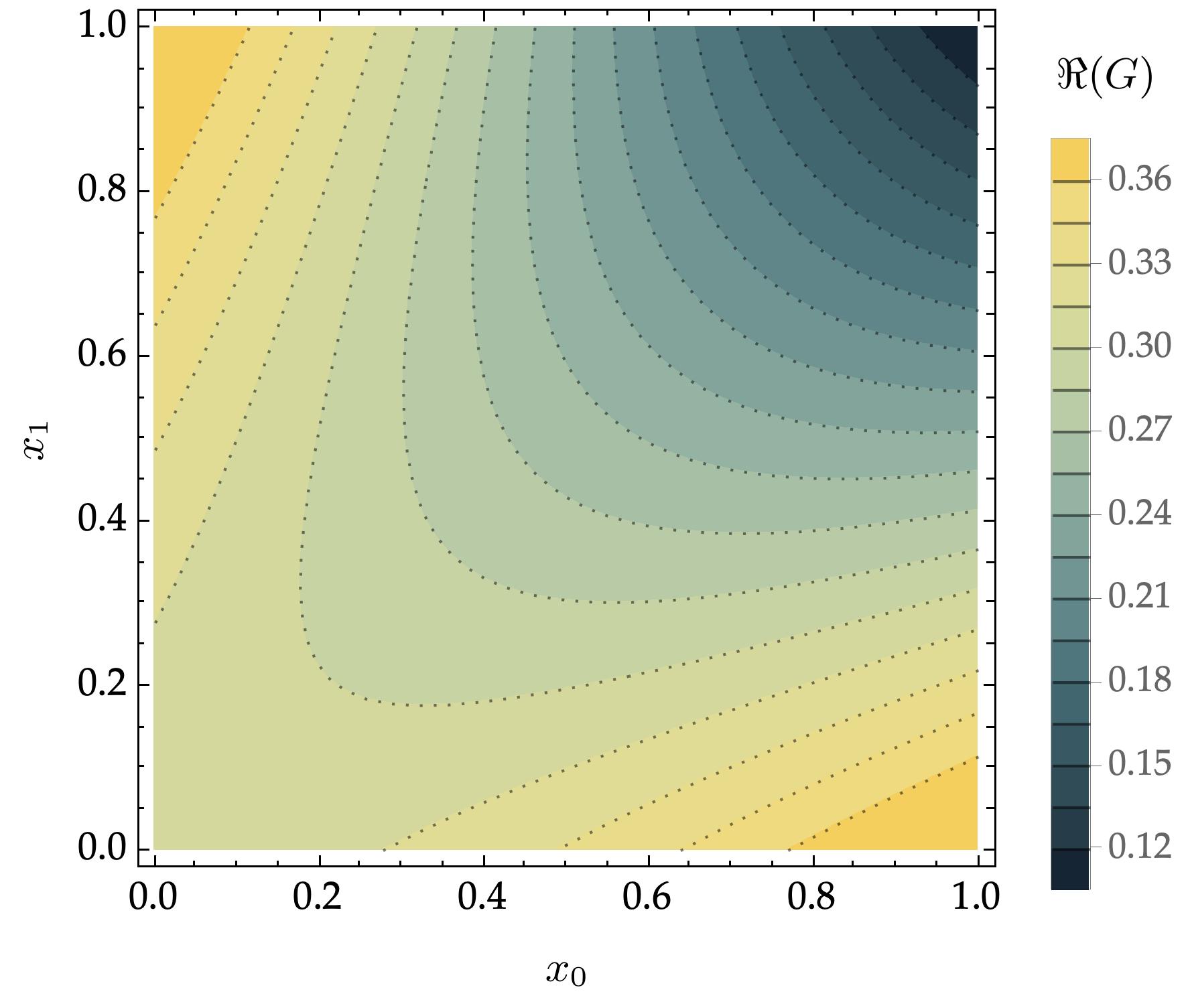}
	\hspace{0.4cm}
	\includegraphics[width=.483\textwidth]{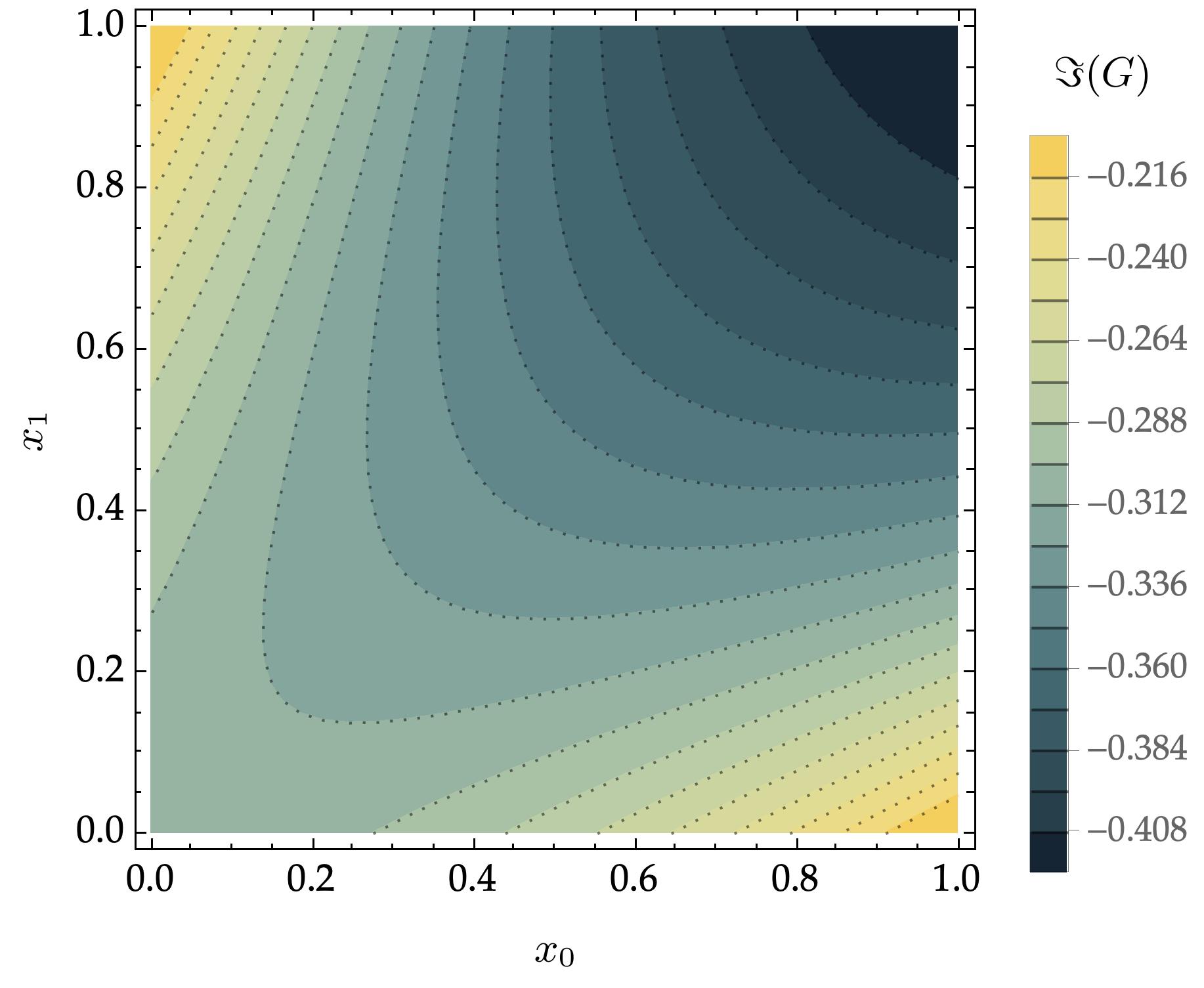}
	\caption{\label{Fig:GParticle} Real and imaginary parts of $G$ as a function of the initial and final positions $x_0$ and $x_1$.}
\end{figure}

In a configuration-space path-integral analysis of the non-relativistic particle, the quantum propagator from $x_0$ at the initial time  $\tau_0 = 0$ to $x_1$ at the final stime $\tau_1 = 1$ is expressed as
\ba\label{eq:PathIntegralParticleNonRel}
G(x_1,1; x_0,0) = \int_{x(0) = x_0}^{x(1)=x_1} \mathcal{D}x(\tau) \,e^{\imath S[x(\tau)]}\,.
\ea
This functional integral can be evaluated as a Gaussian integral. In this case the WKB approximation\,,
\ba\label{eq:GParticle}
G(x_1,1;x_0,0) \,\,\sim \,\, \sum_{\sigma} \mathcal{J}_\sigma\,
e^{\imath
	S_\sigma}\,\,\,\quad \text{where}\quad  \,\,\, \mathcal{J}_\sigma =  \sqrt{\frac{1}{2 \pi \imath }\pdv[2]{S_\sigma}{x_0}{ x_1} }\,,
\ea
is exact. The unique solution to the Euler-Lagrange equations of motion~\eqref{eq:EOMELParticle} satisfying $x(0) = x_0$ and $x(1) = x_1$ is given by
\ba
x(\tau) = x_0 \cos(\tau) + \frac{1}{\sin(1)}\qty(x_1 - x_0 \cos(1))\sin(\tau)\,.
\ea
The action~\eqref{eq:ActionNonRelParticle} onshell evaluates to
\ba
S(x_0,x_1) = \frac{1}{2 \sin(1)}\qty[\qty(x_0^2 + x_1^2) \cos(1) - 2 x_0 x_1] \,.
\ea
Therefrom we can compute the Hessian determinant factor
\ba
\mathcal{J}(x_0,x_1) = \sqrt{\frac{1}{2\pi \imath \sin(1)}} \,.
\ea
Altogether, the saddle-point approximation~\eqref{eq:GParticle} yields
\ba\label{eq:GParticleExplicit}
G(x_1,1;x_0,0) = \sqrt{\frac{1}{2 \pi \imath \sin(1)}} \,e^{ \frac{\imath}{2 \sin(1)}\qty[\qty(x_0^2 + x_1^2) \cos(1) - 2 x_0 x_1]}\,.
\ea
Figure~\ref{Fig:GParticle} shows the real and imaginary parts of $G$ as a function of $x_0$ and $x_1$.

\bibliographystyle{jhep}
\bibliography{references}

@article{Carballo-Rubio:2025fnc,
	author = "Carballo-Rubio, Ra{\'u}l and others",
	title = "{Towards a non-singular paradigm of black hole physics}",
	eprint = "2501.05505",
	archivePrefix = "arXiv",
	primaryClass = "gr-qc",
	doi = "10.1088/1475-7516/2025/05/003",
	journal = "JCAP",
	volume = "05",
	pages = "003",
	year = "2025"
}

@article{Giddings:1988cx,
	author = "Giddings, Steven B. and Strominger, Andrew",
	title = "{Loss of incoherence and determination of coupling constants in quantum gravity}",
	reportNumber = "HUTP-88/A006",
	doi = "10.1016/0550-3213(88)90109-5",
	journal = "Nucl. Phys. B",
	volume = "307",
	pages = "854--866",
	year = "1988"
}

@article{Isichei:2025tvg,
	author = "Isichei, Raymond and Magueijo, Jo{\~a}o",
	title = "{Thermodynamics in space-times without horizons}",
	eprint = "2507.03469",
	archivePrefix = "arXiv",
	primaryClass = "gr-qc",
	month = "7",
	year = "2025"
}

@article{Brown:1992kc,
	author = "Brown, J. David",
	title = "{Action functionals for relativistic perfect fluids}",
	eprint = "gr-qc/9304026",
	archivePrefix = "arXiv",
	reportNumber = "CTMP-001-NCSU",
	doi = "10.1088/0264-9381/10/8/017",
	journal = "Class. Quant. Grav.",
	volume = "10",
	pages = "1579--1606",
	year = "1993"
}

@article{Morris:1988tu,
	author = "Morris, M. S. and Thorne, K. S. and Yurtsever, U.",
	title = "{Wormholes, Time Machines, and the Weak Energy Condition}",
	doi = "10.1103/PhysRevLett.61.1446",
	journal = "Phys. Rev. Lett.",
	volume = "61",
	pages = "1446--1449",
	year = "1988"
}

@article{Borissova:2024hkc,
	author = "Borissova, Johanna and Eichhorn, Astrid and Ray, Shouryya",
	title = "{A non-local way around the no-global-symmetries conjecture in quantum gravity?}",
	eprint = "2407.09595",
	archivePrefix = "arXiv",
	primaryClass = "hep-th",
	doi = "10.1088/1361-6382/ada2d4",
	journal = "Class. Quant. Grav.",
	volume = "42",
	number = "3",
	pages = "037001",
	year = "2025"
}

@article{Geroch:1967fs,
	author = "Geroch, Robert P.",
	title = "{Topology in general relativity}",
	doi = "10.1063/1.1705276",
	journal = "J. Math. Phys.",
	volume = "8",
	pages = "782--786",
	year = "1967"
}

@article{Anderson:1986ww,
	author = "Anderson, A. and DeWitt, Bryce S.",
	title = "{Does the Topology of Space Fluctuate?}",
	doi = "10.1007/BF01889374",
	journal = "Found. Phys.",
	volume = "16",
	pages = "91--105",
	year = "1986"
}

@inproceedings{Dowker:2002hm,
	author = "Dowker, Fay",
	title = "{Topology change in quantum gravity}",
	booktitle = "{Workshop on Conference on the Future of Theoretical Physics and Cosmology in Honor of Steven Hawking's 60th Birthday}",
	eprint = "gr-qc/0206020",
	archivePrefix = "arXiv",
	pages = "436--452",
	month = "6",
	year = "2002"
}

@article{Louko:1995jw,
	author = "Louko, Jorma and Sorkin, Rafael D.",
	title = "{Complex actions in two-dimensional topology change}",
	eprint = "gr-qc/9511023",
	archivePrefix = "arXiv",
	reportNumber = "SU-GP-95-5-1, WISC-MILW-95-TH-16, MDDP-PP-96-40",
	doi = "10.1088/0264-9381/14/1/018",
	journal = "Class. Quant. Grav.",
	volume = "14",
	pages = "179--204",
	year = "1997"
}

@article{Basile:2025zjc,
	author = "Basile, Ivano and Knorr, Benjamin and Platania, Alessia and Schiffer, Marc",
	title = "{Asymptotic safety, quantum gravity, and the swampland: a conceptual assessment}",
	eprint = "2502.12290",
	archivePrefix = "arXiv",
	primaryClass = "hep-th",
	reportNumber = "MPP-2025-37",
	month = "2",
	year = "2025"
}

@article{Buoninfante:2024yth,
	author = "Buoninfante, Luca and others",
	title = "{Visions in quantum gravity}",
	eprint = "2412.08696",
	archivePrefix = "arXiv",
	primaryClass = "hep-th",
	doi = "10.21468/SciPostPhysCommRep.11",
	month = "12",
	year = "2024"
}

@article{Banihashemi:2024aal,
	author = "Banihashemi, Batoul and Jacobson, Ted",
	title = "{On the lapse contour in the gravitational path integral}",
	eprint = "2405.10307",
	archivePrefix = "arXiv",
	primaryClass = "hep-th",
	doi = "10.1103/PhysRevD.111.066014",
	journal = "Phys. Rev. D",
	volume = "111",
	number = "6",
	pages = "066014",
	year = "2025"
}

@article{Hartle:1986yu,
	author = "Hartle, James B. and Kuchar, Karel V.",
	title = "{Path integrals in parametrized theories: The Free relativistic particle}",
	doi = "10.1103/PhysRevD.34.2323",
	journal = "Phys. Rev. D",
	volume = "34",
	pages = "2323--2331",
	year = "1986"
}

@article{Teitelboim:1983fk,
	author = "Teitelboim, Claudio",
	title = "{The Proper Time Gauge in Quantum Theory of Gravitation}",
	reportNumber = "PRINT-83-0135 (TEXAS)",
	doi = "10.1103/PhysRevD.28.297",
	journal = "Phys. Rev. D",
	volume = "28",
	pages = "297",
	year = "1983"
}

@article{Hayward:1990tz,
	author = "Hayward, G. and Louko, J.",
	title = "{Variational principles for nonsmooth metrics}",
	doi = "10.1103/PhysRevD.42.4032",
	journal = "Phys. Rev. D",
	volume = "42",
	pages = "4032--4041",
	year = "1990"
}

@article{Borissova:2025hmj,
	author = "Borissova, Johanna and Liberati, Stefano and Visser, Matt",
	title = "{Timelike convergence condition in regular black-hole spacetimes with (anti-)de Sitter core}",
	eprint = "2509.08590",
	archivePrefix = "arXiv",
	primaryClass = "gr-qc",
	month = "9",
	year = "2025"
}

@article{Dittrich:2024awu,
	author = {Dittrich, Bianca and Jacobson, Ted and Padua-Arg{\"u}elles, Jos{\'e}},
	title = "{de Sitter horizon entropy from a simplicial Lorentzian path integral}",
	eprint = "2403.02119",
	archivePrefix = "arXiv",
	primaryClass = "gr-qc",
	doi = "10.1103/PhysRevD.110.046006",
	journal = "Phys. Rev. D",
	volume = "110",
	number = "4",
	pages = "046006",
	year = "2024"
}

@article{Banihashemi:2024weu,
	author = "Banihashemi, Batoul and Jacobson, Ted",
	title = "{The enigmatic gravitational partition function}",
	eprint = "2411.00267",
	archivePrefix = "arXiv",
	primaryClass = "hep-th",
	doi = "10.1007/s10714-024-03347-0",
	journal = "Gen. Rel. Grav.",
	volume = "57",
	number = "2",
	pages = "43",
	year = "2025"
}

@inproceedings{Horowitz:1991fr,
	author = "Horowitz, Gary T.",
	title = "{Topology change in general relativity}",
	booktitle = "{6th Marcel Grossmann Meeting on General Relativity (MG6)}",
	eprint = "hep-th/9109030",
	archivePrefix = "arXiv",
	reportNumber = "UCSBTH-91-44",
	pages = "1167--1181",
	month = "6",
	year = "1991"
}

@article{Banks:1988yz,
	author = "Banks, Tom and Dixon, Lance J.",
	title = "{Constraints on String Vacua with Space-Time Supersymmetry}",
	reportNumber = "PUPT-1086, SCIPP-8805",
	doi = "10.1016/0550-3213(88)90523-8",
	journal = "Nucl. Phys. B",
	volume = "307",
	pages = "93--108",
	year = "1988"
}

@article{Giddings:1987cg,
	author = "Giddings, Steven B. and Strominger, Andrew",
	title = "{Axion Induced Topology Change in Quantum Gravity and String Theory}",
	reportNumber = "HUTP-87-A067",
	doi = "10.1016/0550-3213(88)90446-4",
	journal = "Nucl. Phys. B",
	volume = "306",
	pages = "890--907",
	year = "1988"
}

@article{Coleman:1989zu,
	author = "Coleman, Sidney R. and Lee, Ki-Myeong",
	title = "{WORMHOLES MADE WITHOUT MASSLESS MATTER FIELDS}",
	reportNumber = "HUTP-89/A022",
	doi = "10.1016/0550-3213(90)90149-8",
	journal = "Nucl. Phys. B",
	volume = "329",
	pages = "387--409",
	year = "1990"
}

@article{Abbott:1989jw,
	author = "Abbott, L. F. and Wise, Mark B.",
	title = "{Wormholes and Global Symmetries}",
	reportNumber = "CALT-68-1523",
	doi = "10.1016/0550-3213(89)90503-8",
	journal = "Nucl. Phys. B",
	volume = "325",
	pages = "687--704",
	year = "1989"
}

@article{Lee:1988ge,
	author = "Lee, Ki-Myeong",
	title = "{Wormholes and Goldstone Bosons}",
	reportNumber = "FERMILAB-PUB-88-027-T",
	doi = "10.1103/PhysRevLett.61.263",
	journal = "Phys. Rev. Lett.",
	volume = "61",
	pages = "263--266",
	year = "1988"
}

@article{Teitelboim:1981ua,
	author = "Teitelboim, Claudio",
	title = "{Quantum Mechanics of the Gravitational Field}",
	reportNumber = "PRINT-81-0842 (IAS,PRINCETON)",
	doi = "10.1103/PhysRevD.25.3159",
	journal = "Phys. Rev. D",
	volume = "25",
	pages = "3159",
	year = "1982"
}

@article{Teitelboim:1983fh,
	author = "Teitelboim, Claudio",
	title = "{Causality Versus Gauge Invariance in Quantum Gravity and Supergravity}",
	reportNumber = "Print-83-0131 (TEXAS)",
	doi = "10.1103/PhysRevLett.50.705",
	journal = "Phys. Rev. Lett.",
	volume = "50",
	pages = "705",
	year = "1983"
}

@article{Halliwell:1988ik,
	author = "Halliwell, Jonathan J. and Louko, Jorma",
	title = "{Steepest Descent Contours in the Path Integral Approach to Quantum Cosmology. 1. The De Sitter Minisuperspace Model}",
	reportNumber = "NSF-ITP-88-173",
	doi = "10.1103/PhysRevD.39.2206",
	journal = "Phys. Rev. D",
	volume = "39",
	pages = "2206",
	year = "1989"
}

@article{Halliwell:1988wc,
	author = "Halliwell, Jonathan J.",
	title = "{Derivation of the Wheeler-De Witt Equation from a Path Integral for Minisuperspace Models}",
	reportNumber = "NSF-ITP-88-25",
	doi = "10.1103/PhysRevD.38.2468",
	journal = "Phys. Rev. D",
	volume = "38",
	pages = "2468",
	year = "1988"
}

@article{Visser:1989am,
	author = "Visser, Matt",
	title = "{Quantum Mechanical Stabilization of Minkowski Signature Wormholes}",
	reportNumber = "LA-UR-89-1745",
	doi = "10.1016/0370-2693(90)91588-3",
	journal = "Phys. Lett. B",
	volume = "242",
	pages = "24--28",
	year = "1990"
}

@article{Visser:1989kh,
	author = "Visser, Matt",
	title = "{Traversable wormholes: Some simple examples}",
	eprint = "0809.0907",
	archivePrefix = "arXiv",
	primaryClass = "gr-qc",
	reportNumber = "LA-UR-89-46",
	doi = "10.1103/PhysRevD.39.3182",
	journal = "Phys. Rev. D",
	volume = "39",
	pages = "3182--3184",
	year = "1989"
}

@article{Visser:1989kg,
	author = "Visser, Matt",
	title = "{Traversable wormholes from surgically modified Schwarzschild space-times}",
	eprint = "0809.0927",
	archivePrefix = "arXiv",
	primaryClass = "gr-qc",
	reportNumber = "LA-UR-89-244",
	doi = "10.1016/0550-3213(89)90100-4",
	journal = "Nucl. Phys. B",
	volume = "328",
	pages = "203--212",
	year = "1989"
}

@article{Visser:1990wi,
	author = "Visser, Matt",
	title = "{Quantum wormholes}",
	reportNumber = "PRINT-90-0514 (WASH.U.,ST.LOUIS)",
	doi = "10.1103/PhysRevD.43.402",
	journal = "Phys. Rev. D",
	volume = "43",
	pages = "402--409",
	year = "1991"
}

@article{Visser:1990wh,
	author = "Visser, Matt",
	title = "{Wheeler wormholes and topology change}",
	reportNumber = "PRINT-90-0515 (WASH-U.,ST.LOUIS)",
	doi = "10.1142/S0217732391003109",
	journal = "Mod. Phys. Lett. A",
	volume = "6",
	pages = "2663--2668",
	year = "1991"
}

@article{Redmount:1993ue,
	author = "Redmount, Ian H. and Suen, Wia-Mo",
	title = "{Quantum dynamics of Lorentzian space-time foam}",
	eprint = "gr-qc/9309017",
	archivePrefix = "arXiv",
	reportNumber = "WUGRAV-93-7",
	doi = "10.1103/PhysRevD.49.5199",
	journal = "Phys. Rev. D",
	volume = "49",
	pages = "5199--5210",
	year = "1994"
}

@book{Visser:1995cc,
	author = "Visser, Matt",
	title = "{Lorentzian wormholes: From Einstein to Hawking}",
	isbn = "978-1-56396-653-8",
	year = "1995"
}

@article{Berry:2020tky,
	author = "Berry, Thomas and Lobo, Francisco S. N. and Simpson, Alex and Visser, Matt",
	title = "{Thin-shell traversable wormhole crafted from a regular black hole with asymptotically Minkowski core}",
	eprint = "2008.07046",
	archivePrefix = "arXiv",
	primaryClass = "gr-qc",
	doi = "10.1103/PhysRevD.102.064054",
	journal = "Phys. Rev. D",
	volume = "102",
	number = "6",
	pages = "064054",
	year = "2020"
}

@article{Borissova:2025msp,
	author = "Borissova, Johanna and Liberati, Stefano and Visser, Matt",
	title = "{Violations of the null convergence condition in kinematical transitions between singular and regular black holes, horizonless compact objects, and bounces}",
	eprint = "2502.00548",
	archivePrefix = "arXiv",
	primaryClass = "gr-qc",
	doi = "10.1103/PhysRevD.111.104054",
	journal = "Phys. Rev. D",
	volume = "111",
	number = "10",
	pages = "104054",
	year = "2025"
}

@article{Hochberg:1998ii,
	author = "Hochberg, David and Visser, Matt",
	title = "{The Null energy condition in dynamic wormholes}",
	eprint = "gr-qc/9802048",
	archivePrefix = "arXiv",
	reportNumber = "LAEFF-98-02",
	doi = "10.1103/PhysRevLett.81.746",
	journal = "Phys. Rev. Lett.",
	volume = "81",
	pages = "746--749",
	year = "1998"
}

@article{Simpson:2019cer,
	author = "Simpson, Alex and Martin-Moruno, Prado and Visser, Matt",
	title = "{Vaidya spacetimes, black-bounces, and traversable wormholes}",
	eprint = "1902.04232",
	archivePrefix = "arXiv",
	primaryClass = "gr-qc",
	doi = "10.1088/1361-6382/ab28a5",
	journal = "Class. Quant. Grav.",
	volume = "36",
	number = "14",
	pages = "145007",
	year = "2019"
}

@article{Simpson:2018tsi,
	author = "Simpson, Alex and Visser, Matt",
	title = "{Black-bounce to traversable wormhole}",
	eprint = "1812.07114",
	archivePrefix = "arXiv",
	primaryClass = "gr-qc",
	doi = "10.1088/1475-7516/2019/02/042",
	journal = "JCAP",
	volume = "02",
	pages = "042",
	year = "2019"
}

@article{Morris:1988cz,
	author = "Morris, M. S. and Thorne, K. S.",
	title = "{Wormholes in space-time and their use for interstellar travel: A tool for teaching general relativity}",
	doi = "10.1119/1.15620",
	journal = "Am. J. Phys.",
	volume = "56",
	pages = "395--412",
	year = "1988"
}

\end{document}